\documentclass[prd,amsmath,amssymb,nofootinbib,superscriptaddress,showkeys,reprint]{revtex4-2}
\usepackage{graphicx}
\usepackage{booktabs}
\usepackage{array}
\usepackage{multirow}
\usepackage{dcolumn}
\usepackage{bm}
\usepackage{cancel}

\usepackage[caption=false]{subfig}
\usepackage[table]{xcolor}
\definecolor{MAGENTA}{named}{magenta}
\usepackage[colorlinks=true, linkcolor=blue, citecolor=blue, urlcolor=blue]{hyperref}
\usepackage{orcidlink}
\usepackage{siunitx}

\newcommand{\dd}{\mathrm{d}}

\newcolumntype{B}{>{\bfseries}c}

\begin{document}

\title{Thermodynamic geometry of charged AdS black holes with a string cloud in Lorentz-violating Einstein–Kalb–Ramond gravity}

\author{Faizuddin Ahmed\orcidlink{0000-0003-2196-9622}} 
\email[Faizuddin Ahmed - ]{faizuddinahmed15@gmail.com} 
\affiliation{Department of Physics, The Assam Royal Global University, Guwahati, 781035, Assam, India}

\author{Edilberto O. Silva\orcidlink{0000-0002-0297-5747}}
\email[Edilberto O. Silva - ]{edilberto.silva@ufma.br (Corresp. author)}
\affiliation{Programa de P\'os-Gradua\c c\~ao em F\'{\i}sica \& Coordena\c c\~ao do Curso de F\'{\i}sica -- Bacharelado, Universidade Federal do Maranh\~{a}o, 65085-580 S\~{a}o Lu\'{\i}s, Maranh\~{a}o, Brazil}

\date{\today}

\begin{abstract}
We investigate the thermodynamic microstructure of electrically charged AdS black holes in Einstein--Kalb--Ramond bumblebee gravity in the presence of a spherically symmetric cloud of strings. Employing Weinhold and Ruppeiner thermodynamic geometries in complementary thermodynamic representations, we show that curvature singularities consistently track the spinodal boundaries and the second-order critical point associated with a Van der Waals--like small/large black-hole phase transition. Moreover, the sign structure of the Ruppeiner curvature provides a transparent characterization of the competing microscopic interactions across the coexisting phases. We find that the Lorentz-violating parameter $\ell$ and the string-cloud parameter $\alpha$ shift the critical scales and rescale correlation measures while preserving the universality class of the critical behavior. We further comment on dynamical and holographic implications and contrast the thermodynamic sensitivity to $(\ell,\alpha)$ with thin-disk optical signatures.
\end{abstract}

\maketitle

\section{Introduction}
\label{sec:intro}

Black hole thermodynamics has long provided a privileged arena to test fundamental aspects of gravity, quantum theory, and statistical mechanics. Since the formulation of the four laws of black hole mechanics and the identification of entropy and temperature with geometric quantities on the horizon~\cite{PRD.1973.7.2333,CMP.1973.31.161,CMP.1975.43.199}, black holes have been regarded as genuine thermodynamic systems. The development of the extended phase space approach, in which the cosmological constant is promoted to a thermodynamic pressure and the black hole mass is interpreted as enthalpy~\cite{CQG.2009.26.195011,CQG.2011.28.125020,PRD.2011.84.024037}, has further enriched this picture by revealing a broad spectrum of phase transitions analogous to those of ordinary fluids. In particular, charged AdS black holes display a small/large black hole phase transition that closely mimics the liquid--gas transition of the Van der Waals fluid~\cite{Chamblin.1999.60.064018,Chamblin.1999.60.104026,JHEP.2012.2012.033}, with critical exponents consistent with mean--field theory and a universal ratio $P_c v_c/T_c=3/8$.

In recent years, this thermodynamic viewpoint has been extended to a wide variety of gravitational backgrounds and matter contents, including higher--curvature theories, nonlinear electrodynamics, massive gravity, Horndeski and Lovelock models, among others~\cite{JHEP.2013.2013.005,PRD.2013.87.044014,JHEP.2012.2012.110,PRD.2013.88.084045,EPJC.2020.80.17,PDU.2020.30.100730}. These generalizations show that the Van der Waals--like structure is remarkably robust: modified gravity and deformed matter sectors typically change the location of the critical point and the detailed phase diagram, but the universality class often remains mean--field. At the same time, a complementary thermodynamic formalism based on Joule--Thomson expansion, heat engines, and irreversible processes has been developed~\cite{PRD.2018.98.124032,PRD.2018.98.084014,EPJC.2017.77.24,EPJC.2018.78.123,PLB.2019.797.134883,FrontPhys.2021.9.628727}, providing additional probes of the underlying microscopic structure beyond equilibrium thermodynamics.

Thermodynamic geometry offers a powerful, intrinsically geometric language to encode this information. In Weinhold's formulation, the state space is endowed with a Riemannian metric defined as the Hessian of the internal energy in the energy representation~\cite{JCP.1975.63.2479,JCP.1975.63.2484}. Ruppeiner's construction instead uses the negative Hessian of the entropy in an appropriate representation, leading to a metric tensor that is directly linked to the fluctuation theory of equilibrium thermodynamics and information geometry~\cite{PRA.1979.20.1608,RMP.1995.67.605,SPP.2014.153.179}. For black holes, both approaches have been widely applied to study stability, criticality, and, more recently, to extract hints about the microscopic interactions that might underlie black hole thermodynamics~\cite{GRG.2003.35.1733,JMP.2007.48.013506,EPL.2012.99.20004,PRL.2015.115.111302,AHEP.2017.2017.7158697,PRD.2019.99.044013}. In this context, the Ruppeiner scalar curvature $R_{\mathrm{R}}$ is often interpreted as a measure of the effective correlation volume: $R_{\mathrm{R}}=0$ corresponds to an ideal gas; $R_{\mathrm{R}}<0$ is associated with effectively attractive, Van der Waals--type interactions; and $R_{\mathrm{R}}>0$ signals predominantly repulsive, Fermi--gas--like behavior.

The interplay between thermodynamic geometry and phase structure has been explored for a variety of AdS black holes, including Reissner--Nordström, rotating, Gauss--Bonnet, Born--Infeld, and massive gravity solutions~\cite{PRD.2013.87.044014,PRD.2013.88.084045,PRD.2017.95.021501,EPJC.2020.80.17,JHEP.2015.2015.143}. A robust picture has emerged: the divergences of $R_{\mathrm{R}}$ typically coincide with singularities of the specific heat or isothermal compressibility, thereby tracking second--order critical points and spinodal curves, while the sign of $R_{\mathrm{R}}$ distinguishes different thermodynamic phases and interaction regimes. More recently, this geometric viewpoint has been complemented by a topological classification of black hole phase structures based on the Duan $\phi$--mapping and winding numbers in an auxiliary parameter space~\cite{Mansoori2014,Mansoori2015,PRD.2018.97.104027,PRL.2022.129.191101,PRD.2022.105.104003,PRD.2022.105.104053,PRD.2023.107.064015,EPJC.2023.83.944,EPJC.2023.83.957,PDU.2025.102079}. In this approach, phase transitions and critical phenomena are associated with topological defects in the thermodynamic parameter manifold, whose total topological charge is remarkably robust under deformations of the underlying theory.

Parallel to these developments, there has been sustained interest in black hole solutions coupled to nontrivial matter distributions such as clouds of strings, which model ensembles of one-dimensional topological defects~\cite{PRD.1979.20.1294}. Clouds of strings can arise as remnants of cosmological phase transitions or as effective descriptions of fundamental strings in certain regimes, and their gravitational effects have been studied in a wide range of contexts, including Schwarzschild--like, Reissner--Nordström, and rotating black holes~\cite{PLB.2018.785.105,EPJC.2018.78.534,EPJC.2019.79.117,CPC.2018.42.063105,Universe.2024.10.430,EPJC.2020.80.17,EPJC.2022.82.227,PDU.2020.30.100730,MPLA.2024.39.232350181X,MPLA.2023.38.2350102,MPLA.2024.39.2450164}. In the AdS context, string clouds modify the thermodynamic variables, shift the critical point, and can generate novel interplay between topological defects and black hole chemistry, with potential implications for holographic duals in strongly coupled condensed matter systems~\cite{PLB.2018.785.105,EPJC.2018.78.534,EPJC.2019.79.117,Universe.2024.10.430,MPLA.2024.39.232350181X,MPLA.2023.38.2350102,MPLA.2024.39.2450164}. In particular, the presence of a cloud of strings has been shown to affect phase transitions, transport coefficients, and KSS bounds in Einstein--Gauss--Bonnet and Rastall gravities, as well as in Yang--Mills charged AdS backgrounds~\cite{MPLA.2024.39.232350181X,MPLA.2023.38.2350102,MPLA.2024.39.2450164}.

Another key ingredient of the present work is Lorentz violation (LV), which has attracted considerable attention as a possible low-energy remnant of Planck-scale physics. In effective field theory, Lorentz violation can be systematically described within the Standard Model Extension (SME)~\cite{PRD.2004.69.105009,PRD.2008.77.125007} or via vector--tensor theories in which a vector field acquires a nonzero vacuum expectation value, thereby spontaneously breaking local Lorentz symmetry~\cite{PRD.2010.82.125015,PRD.2010.81.065028}. Bumblebee gravity and related LV models have led to a rich black hole phenomenology, including modified horizon structures, shadows, lensing properties, and quasi--normal modes~\cite{EPJC.2020.80.335,PRD.2023.108.124004,EPJC.2024.84.964,EPJC.2025.85.658,PRD.2024.110.024077,CQG.2025.42.065026,PDU.2025.102079,AraujoFilho2025NonmetricityBumblebee}. In addition, Lorentz--violating Kalb--Ramond sectors have recently been analyzed from the combined viewpoint of thermodynamics, geometry, and particle dynamics~\cite{PDU.2025.50.102086}, further reinforcing the interplay between LV backgrounds and black hole microphysics.

When combined with antisymmetric tensor fields such as the Kalb--Ramond (KR) field, which naturally arises in string theory and can mimic axion--like or torsion--like degrees of freedom~\cite{EPJC.2020.80.335,PRD.2023.108.124004,EPJC.2024.84.964}, these LV sectors lead to further structure in the black hole solutions, including modified charges, effective topological couplings, and nontrivial matter profiles. Kalb--Ramond black holes and related antisymmetric--tensor deformations have been investigated in detail in connection with particle creation and evaporation, noncommutative geometries, geodesic motion, deflection angles, scattering, quasinormal modes, and black hole shadows~\cite{AraujoFilho2025KRParticleCreation,AraujoFilho2025NoncommKR,AraujoFilho2025ParticleMotionKR,AraujoFilho2025ImpactAntisymmetric,AraujoFilho2024AntisymmetricShadows,PDU.2025.50.102086}, providing a broad arena to test how antisymmetric fields and LV parameters affect both thermodynamic and observational properties.

The specific combination of the Kalb--Ramond bumblebee sector with a cloud of strings is physically motivated by several compelling theoretical scenarios. The Lorentz-violating parameter $\ell$ naturally emerges in contexts of quantum gravity where Lorentz symmetry is spontaneously broken at high energies, providing a low-energy effective description of such fundamental physics. Simultaneously, the string cloud parameter $\alpha$ models the gravitational influence of a distribution of one-dimensional topological defects, which could be remnants of phase transitions in the early universe or fundamental strings in a cosmological context. The synergy between these elements allows us to investigate a pressing question: how do extended material defects interact with spacetime geometries where the fundamental Lorentz symmetry is broken? This investigation is not merely mathematical but has potential implications for cosmological models and for understanding the robustness of black hole thermodynamics in the presence of Lorentz-violating backgrounds and matter sources~\cite{PLB.2018.785.105,EPJC.2018.78.534,EPJC.2019.79.117,Universe.2024.10.430,EPJC.2020.80.335,PRD.2023.108.124004,EPJC.2024.84.964,PDU.2025.50.102086,AraujoFilho2025KRParticleCreation,AraujoFilho2025ParticleMotionKR,AraujoFilho2025NonmetricityBumblebee}.

In a recent work Ref.~\cite{Ahmed2025}, we constructed and analyzed an electrically charged AdS black hole solution in Kalb--Ramond bumblebee gravity sourced by a cloud of strings (EKR black hole). Working in the extended phase space, we showed that the presence of the LV and string cloud parameters $(\ell,\alpha)$ preserves the Van der Waals-like critical behavior of the Reissner--Nordstr\"om-AdS black hole while deforming the location of the critical point and the detailed $P P$-$V$diagram. We further investigated the phase structure using the thermodynamic topology approach~\cite{PRD.2018.97.104027,PRL.2022.129.191101,PRD.2022.105.104003}, finding that the total topological charge remains $W=1$, as in the standard RN--AdS case, and that the deformation parameters shift but do not change the number or type of thermodynamic defects. These results suggest that the EKR black hole provides a robust, ttwo-parameterfamily of deformations of the RN--AdS thermodynamic landscape, in which Lorentz violation and string clouds act as controllable ``knobs'' without altering the underlying universality class.

The natural next step, pursued in the present work, is to investigate the \emph{microscopic} interpretation of this deformed phase structure via thermodynamic geometry. By constructing the Weinhold and Ruppeiner metrics directly from the exact internal energy and Hawking temperature of the EKR black hole, we analyze how the scalar curvatures $R_{\mathrm{W}}$ and $R_{\mathrm{R}}$ encode the influence of $(\ell,\alpha)$ on thermodynamic correlations, critical behavior, and phase coexistence. In particular, we show that the curvature singularities of both metrics coincide with the divergences of the specific heat and the critical point identified in the $P$--$V$ diagram, thereby confirming the consistency between geometric and standard thermodynamic diagnostics. Moreover, the sign structure of $R_{\mathrm{R}}$ in the $(S,Q)$ plane reveals a clear separation between small--black--hole phases with effectively repulsive (Fermi-like) microscopic behavior and large black-hole phases with effectively attractive (Van der Waals-like) interactions, with the transition region and curvature magnitude being tunable by $(\ell,\alpha)$.

Our analysis also connects the geometric description to the thermodynamic topology of Ref.~\cite{Ahmed2025}: the Ruppeiner curvature singularities trace the boundaries separating regions with different winding numbers in the topological classification, while the total topological charge remains invariant under deformations of $(\ell,\alpha)$~\cite{Mansoori2014,Mansoori2015,PRD.2018.97.104027,PRL.2022.129.191101}. Finally, we discuss how dynamical probes such as quasi--normal modes and relaxation times, which have recently been studied in LV and Kalb--Ramond black holes~\cite{PRD.2018.97.104027,EPJC.2025.85.658,PDU.2025.102079,AraujoFilho2025ImpactAntisymmetric,AraujoFilho2024AntisymmetricShadows,AraujoFilho2025KRParticleCreation,AraujoFilho2025ParticleMotionKR}, might be correlated with the Ruppeiner curvature landscape of the EKR solution, opening a route to connect thermodynamic microstructure, LV signatures, and ringdown observables in AdS backgrounds.

The paper is organized as follows. In Sec.~\ref{sec:review}, we briefly review the EKR black hole solution, its thermodynamic variables, and the main results of Ref. \cite{Ahmed2025} regarding extended phase space thermodynamics and topological classification. In Sec.~\ref{sec:geometry}, we construct the Weinhold and Ruppeiner thermodynamic geometries in the $(S,Q)$ state space and analyze their curvature properties, highlighting the role of the LV and string cloud parameters. Sec.~\ref{sec:critical-geometry} is devoted to the critical behavior and correlation lengths: we study the scaling of the Ruppeiner curvature near the critical point, its geometric signatures of phase coexistence, and its interplay with thermodynamic topology. In Sec.~\ref {sec:dynamical}, we discuss dynamical and topological probes, emphasizing possible connections between thermodynamic geometry, quasi-normal modes, and relaxation times in the presence of Lorentz violation and string clouds. We conclude in Sec.~\ref{sec:conclusions} with a summary of our main results and perspectives for future work.

\section{Review of the EKR black hole with a cloud of strings}
\label{sec:review}

\subsection{Metric, parameters and horizon structure}

In this subsection, we briefly summarize the charged AdS black hole solution of Einstein--Kalb--Ramond (EKR) bumblebee gravity in the presence of a spherically symmetric cloud of strings, which will provide the geometric background for our thermodynamic-geometry analysis. Throughout this work, we follow the conventions and notation of our previous thermodynamic study of the same system (see Ref. \cite{Ahmed2025}), and we work in units where $G=c=\hbar=k_B=1$.

The Lorentz-violating (LV) sector of the theory is induced by a background Kalb--Ramond field whose vacuum expectation value deforms the Einstein equations via an effective dimensionless parameter
\begin{equation}
    \ell = \frac{\xi\,b^2}{2},\label{aa1}
\end{equation}
where $\xi$ is the LV coupling and $b$ is the vacuum expectation value of the KR field. A nonvanishing $\ell$ encodes spontaneous breaking of local Lorentz symmetry and leads to effective rescalings of several metric coefficients. The matter sector is modeled by a spherically symmetric cloud of strings, characterized by a stress-energy tensor of the form $T^{t}{}_{t}=T^{r}{}_{r}=\alpha/r^2$, with all angular components vanishing. The dimensionless parameter $\alpha$ controls the solid-angle deficit and the effective mass distribution associated with the cloud; in the limit $\alpha\to 0$ the string cloud is switched off and one recovers the usual EKR bumblebee black hole.

Combining the LV and string-cloud contributions, the static and spherically symmetric charged AdS solution can be written as
\begin{equation}
    ds^2 = -f(r)\,dt^2 + \frac{dr^2}{f(r)} + r^2\left(d\theta^2+\sin^2\theta\,d\phi^2\right),\label{metric}
\end{equation}
with lapse function
\begin{equation}
f(r) =
\frac{1-\alpha}{1-\ell}
- \frac{2M}{r}
+ \frac{Q_\mathrm{eff}^{\,2}}{r^2}
- \frac{\Lambda_\mathrm{eff}}{3}\,r^2,
\label{eq:f_metric}
\end{equation}
where the effective charge and cosmological constant are
\begin{equation}
    Q_\mathrm{eff} = \frac{Q}{1-\ell}, 
    \qquad
    \Lambda_\mathrm{eff} = \frac{\Lambda}{1-\ell}.\label{aa2}
\end{equation}
Here $M$ is the ADM mass, $Q$ is the physical electric charge, and $\Lambda<0$ is the bare cosmological constant; the AdS curvature scale is related to $\Lambda_\mathrm{eff}$ in the usual way. The LV parameter $\ell$ and the string-cloud parameter $\alpha$ thus enter the geometry primarily by rescaling the constant term and dressing the mass and charge contributions in $f(r)$.

The event horizon $r_+$ is defined as the largest positive root of $f(r)=0$. In the absence of a cosmological constant, $\Lambda=0$, the polynomial structure of $f(r)$ allows one to write the inner and outer horizon radii $r_\pm$ in closed form, revealing an extremality condition on the charge-to-mass ratio,
\begin{equation}
\frac{Q^2}{M^2} \le (1-\ell)^3(1-\alpha)^{-1},\label{aa3}
\end{equation}
which generalizes the Reissner--Nordström bound to include both Lorentz violation and the string cloud. For nonzero $\Lambda<0$, the AdS term $-\Lambda_\mathrm{eff} r^2/3$ dominates at large $r$, ensuring the presence of a single event horizon for the parameter choices considered in Ref. \cite{Ahmed2025}.

It is convenient to regard the parameters $(M,Q,\Lambda;\ell,\alpha)$ as specifying a five-dimensional family of black holes: $(M,Q,\Lambda)$ control the usual mass, charge and AdS curvature, while $(\ell,\alpha)$ play the role of deformation parameters that encode the backreaction of the KR field and the string cloud. The thermodynamic analysis in Ref. \cite{Ahmed2025} shows that $(\ell,\alpha)$ can be tuned independently to deform the horizon structure, the extremal limit, and the phase diagram, without changing the basic Reissner-Nordström-AdS character of the solution.

\subsection{Extended phase space and criticality: summary of previous results}

We now summarize the main thermodynamic properties and critical behavior of the charged EKR--string AdS black hole obtained in Ref. \cite{Ahmed2025}. These results will be used as input for the thermodynamic-geometry analysis developed in the subsequent sections.

In the extended phase-space formalism, the cosmological constant is interpreted as a thermodynamic pressure. In our LV setup, the relation between $\Lambda$ and $P$ is modified by the factor $(1-\ell)$,
\begin{equation}
    \Lambda = -8\pi(1-\ell)\,P,\label{eq:bb1}
\end{equation}
so that the ADM mass $M$ is identified with the system's enthalpy rather than its internal energy. Evaluating the horizon condition $f(r_+)=0$ and solving for $M$ in terms of the largest root $r_+$, one obtains
\begin{equation}
    M(r_+,Q,P,\ell,\alpha)= \frac{r_+}{2}\left[\frac{1-\alpha}{1-\ell}+ \frac{Q^2}{(1-\ell)^2 r_+^2}+ \frac{8\pi}{3}P r_+^2\right].\label{mass-1}
\end{equation}
This expression reduces to the enthalpy of the charged EKR AdS black hole when $\alpha=0$, and to the enthalpy of a cloud-of-strings AdS black hole without Lorentz violation when $\ell=0$.

The uncorrected entropy follows the Bekenstein-Hawking area law,
\begin{equation}
S = \frac{A}{4} = \pi r_+^2,\label{entropy-1}
\end{equation}
since the LV and string-cloud sectors do not introduce higher-curvature terms in the action. The thermodynamic volume is obtained from the standard definition $V=(\partial M/\partial P)_{S,Q,\ell,\alpha}$, leading to
\begin{equation}
    V = \frac{4\pi}{3}\,r_+^3,\label{volume-1}
\end{equation}
which coincides with the geometric volume enclosed by the horizon, as in Schwarzschild and Reissner--Nordström AdS black holes.

The Hawking temperature can be computed either from the surface gravity or by differentiating the enthalpy with respect to entropy at fixed $(P,Q,\ell,\alpha)$. In either case, one finds
\begin{equation}
   T= \frac{1}{4\pi r_+}\left[\frac{1-\alpha}{1-\ell}- \frac{Q^2}{(1-\ell)^2 r_+^2}+ 8\pi P\,r_+^2\right].\label{temperature}
\end{equation}
The first term encodes the effective solid-angle deficit generated by the combined LV and string-cloud sectors, the second is the usual Coulombic contribution (dressed by $(1-\ell)$), and the third is the AdS contribution proportional to the pressure $P$. For fixed $(Q,P)$, varying $\ell$ and $\alpha$ shifts the zeros and extrema of $T_H(r_+)$, thereby deforming the separation between small and large black hole branches.

With $M$, $S$, $V$, and $T$ at hand, the Gibbs free energy in the extended phase space is defined as
\begin{equation}
    G = M -TS= \frac{r_+}{4}\left[\frac{1-\alpha}{1-\ell}+ \frac{3Q^2}{(1-\ell)^2 r_+^2}- \frac{8\pi}{3}\,P\,r_+^2\right],\label{gibbs-energy-1}
\end{equation}
while the internal energy $U$ is given by
\begin{equation}
    U = M - PV= \frac{r_+}{2}\left[\frac{1-\alpha}{1-\ell}+ \frac{Q^2}{(1-\ell)^2 r_+^2}\right].\label{internal-energy-1}
\end{equation}
Both quantities are deformed relative to the Reissner-Nordström-AdS case by the presence of $(\ell,\alpha)$, which effectively ``dresses'' the constant and Coulombic contributions.

The specific heat at constant pressure is defined as
\begin{equation}
    C_{P} =T\left(\frac{\partial S}{\partial T}\right)_{P,Q,\ell,\alpha}=T\frac{(\partial S/\partial r_+)}{(\partial T/\partial r_+)},\label{definition}
\end{equation}
and, using Eq.~\eqref{temperature}, can be written explicitly as
\begin{equation}
C_{P}= 2\pi r_+^2\frac{\displaystyle \frac{1-\alpha}{1-\ell}- \frac{Q^2}{(1-\ell)^2 r_+^2}+ 8\pi P\,r_+^2}{\displaystyle -\frac{1-\alpha}{1-\ell}+ \frac{3Q^2}{(1-\ell)^2 r_+^2}
+ 8\pi P\,r_+^2}.\label{heat-capacity-1}
\end{equation}
The zeros and divergences of $C_{P}$ delimit locally stable ($C_{P}>0$) and unstable ($C_{P}<0$) branches. For suitable values of $(\ell,\alpha,Q,P)$, reproduce the characteristic three-branch pattern (small, intermediate, and large black holes) associated with Van der Waals-like behavior.

The equation of state is obtained by inverting Eq.~\eqref{temperature} to express the pressure as a function of $(T,\,r_+)$. Writing
\begin{equation}
    P = \frac{T}{2r_+} - \frac{\eta}{4r_+^2} + \frac{\zeta}{16 r_+^4},
    \qquad
    v = 2r_+, \label{EOS}
\end{equation}
with
\begin{equation}
    \eta = \frac{1}{2\pi}\frac{1-\alpha}{1-\ell},
    \qquad
    \zeta = \frac{2Q^2}{\pi(1-\ell)^2},
    \label{eq:review_etazeta}
\end{equation}
one finds that the equation of state $P(T,v)$ has exactly the same functional structure as in the Reissner-Nordström-AdS case, with $\eta$ and $\zeta$ encoding the deformations due to the LV parameter and the string cloud. Imposing the criticality conditions $(\partial P/\partial v)_{T}=0$ and $(\partial^2 P/\partial v^2)_{T}=0$ yields analytic expressions for the critical specific volume $v_c$, critical temperature $T_c$ and critical pressure $P_c$,
\begin{equation}
\left\{
\begin{aligned}
    v_c &= 2\sqrt{\frac{6}{(1-\ell)(1-\alpha)}}\,Q,\\
    T_c &= \frac{(1-\alpha)^{3/2}}{3\pi\sqrt{6}\,Q\sqrt{1-\ell}},\\
    P_c &= \frac{(1-\alpha)^2}{96\pi Q^2}
\end{aligned}
\right.
\label{eq:review_critical}
\end{equation}

and the corresponding universal ratio
\begin{equation}
    \frac{P_c v_c}{T_c} = \frac{3}{8},\label{eq:review_ratio}
\end{equation}
identical to that of Van der Waals fluids and Reissner--Nordström--AdS black holes. Thus, the LV parameter $\ell$ and the string-cloud parameter $\alpha$ deform the critical scales $(v_c,T_c,P_c)$ but preserve the thermodynamic universality class.
\begin{figure*}[t]
\centering
\includegraphics[width=0.98\textwidth]{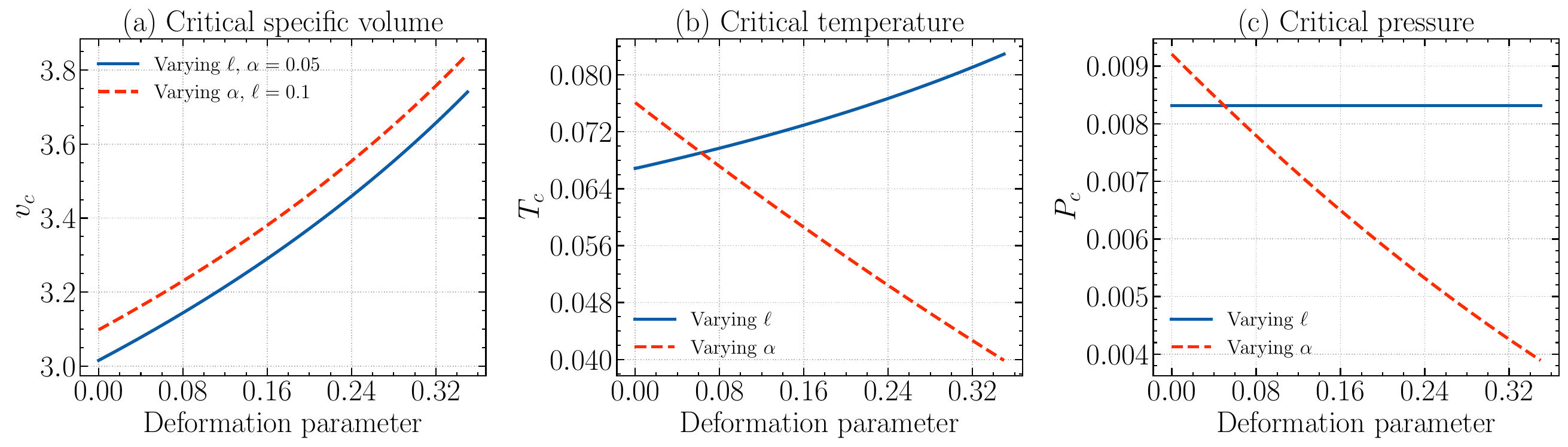}
\caption{(Color online) Critical thermodynamic quantities ($v_c$, $T_c$, and $P_c$) of the charged AdS black hole with a cloud of strings in Kalb-Ramond gravity, obtained from the criticality condition in Eq.~(\ref{eq:review_critical}). We fix $Q = 0.6$ and show the dependence of the critical specific volume $v_c$ (a), the critical temperature $T_c$ (b), and the critical pressure $P_c$ (c) on the deformation parameters. Solid blue curves correspond to varying $\ell$ at fixed $\alpha = 0.05$, whereas dashed red curves correspond to varying $\alpha$ at fixed $\ell = 0.10$.}
\label{fig:critical_vs_deformation}
\end{figure*}
The combined effect of the Lorentz--violating background and the string cloud on the critical point derived in Eq.~(\ref{eq:review_critical}) is summarised in Fig.~\ref{fig:critical_vs_deformation}. Figure \ref{fig:critical_vs_deformation}(a) shows that the critical specific volume $v_c$ increases monotonically with both $\ell$ and $\alpha$, indicating that the onset of criticality requires larger horizon radii in the deformed background. The critical temperature $T_c$ exhibits a more intricate response [Figure \ref{fig:critical_vs_deformation}(b)]: it rises with the string parameter $\ell$. Still, it decreases with the Lorentz--violating parameter $\alpha$, so that the two deformations have competing effects. Finally, Fig. \ref{fig:critical_vs_deformation}(c) confirms that the critical pressure $P_c$ is independent of $\ell$ but decreases linearly with $\alpha$. Altogether, these trends demonstrate that the Lorentz--violating sector tends to lower both the critical temperature and pressure, while the string cloud primarily shifts the critical volume and temperature, reshaping the location of the Van der Waals–like critical point in the extended phase space.

In Ref. \cite{Ahmed2025}, the global phase structure was further characterized by the thermodynamic-topology method of Wei \emph{et al.}, based on a generalized off-shell free energy $\mathcal{F}(r_+,\tau)$ and an associated vector field in an auxiliary $(r_+,\theta)$ space. The zero points of this vector field correspond to black hole phases, and their winding numbers define a total topological charge $W$. For the EKR black hole with a cloud of strings, it was found that $W=1$, the same value as in the Reissner-Nordström-AdS case, indicating that the LV and string-cloud sectors do not change the global topological class of the solution, but only shift the locations and stability properties of individual branches.

Finally, the Joule--Thomson expansion was analyzed by computing the inversion temperature $T_i(P)$ and the corresponding minimal inversion temperature $T_i^{\rm min}$ along isenthalpic curves $M=\mathrm{const.}$ in the $T$--$P$ plane. An important result is that the ratio between the minimal inversion temperature and the critical temperature satisfies
\begin{align}
\frac{T_i^{\rm min}}{T_c} = \frac{1}{2},\label{eq:review_TiTc}
\end{align}
exactly as in charged AdS black holes in Einstein gravity. This shows that, even though $(\ell,\alpha)$ provide new knobs to tune the absolute scales of $T_i$ and $T_c$, the qualitative structure of the heating/cooling regions in the $T$--$P$ diagram remains universal.

In the remainder of this work we will use the thermodynamic relations summarized above as the starting point to construct Ruppeiner and Weinhold metrics, analyze their scalar curvatures, and probe the microscopic structure of the EKR black hole with a string cloud from a geometric viewpoint.

\section{Thermodynamic fluctuations: second-order corrections to entropy}
\label{sec:fluctuations}

In this section, we extend the analysis of the thermodynamic structure of the charged AdS black hole in Kalb-Ramond bumblebee gravity with a cloud of strings by incorporating thermal fluctuations that arise from the statistical interpretation of black hole entropy. While the leading-order Bekenstein-Hawking entropy $S_0 = \pi r_+^2$ [Eq.~\eqref{entropy-1}] provides an excellent approximation for macroscopic black holes, quantum and statistical corrections become increasingly important as the horizon radius decreases and thermal fluctuations intensify~\cite{Das2001,Chatterjee2015,Faizal2015}. Following the seminal framework of Hawking and Page~\cite{HawkingPage}, we treat the black hole as a canonical ensemble characterized by an energy spectrum, and employ a systematic expansion of the entropy to quantify the impact of fluctuations on the thermodynamic geometry and phase structure discussed in the previous sections.

\subsection{Statistical foundation: density of states and partition function}

The statistical partition function for a black hole in thermal equilibrium with a heat reservoir at inverse temperature $\beta_0 = 1/(k_B T)$ (where $k_B$ is Boltzmann's constant, set to unity in our units) can be expressed as a sum over all accessible microstates weighted by their Boltzmann factors. In the continuum limit, this takes the form of a Laplace transform over the energy spectrum~\cite{Gibbons1977, Iyer1995, Ashraf2025}:
\begin{align} 
Z = \int_0^\infty dE \, \rho(E) \, e^{-\beta_0 E},\label{eq:partition_function}
\end{align}
where $\rho(E)$ denotes the microcanonical density of states at energy $E$. The integration extends over all energies consistent with the existence of a black hole horizon; in our case, this corresponds to configurations satisfying $f(r_+) = 0$ [Eq.~\eqref{eq:f_metric}] with $r_+ > 0$.

The key step in connecting the partition function to entropy corrections is to invert the Laplace transform in Eq.~\eqref{eq:partition_function} and express $\rho(E)$ in terms of the thermodynamic entropy $S(E)$. For a system in thermal equilibrium, the density of states can be approximated by a Gaussian distribution centered at the mean energy $\langle E \rangle = E_0$, with variance determined by the second derivative of the entropy with respect to the inverse temperature~\cite{Benham2015,Mushtaq2025}. Specifically, the saddle-point approximation to the inverse Laplace transform yields
\begin{equation}
\rho(E) = \frac{e^{S_0}}{\sqrt{2\pi}} \left[ \left( \frac{\partial^2 S(\beta)}{\partial \beta^2} \right)_{\beta = \beta_0} \right]^{-1/2},\label{eq:density_of_states}
\end{equation}
where $S_0 = S(E_0)$ is the entropy evaluated at the equilibrium energy $E_0$, and the derivative is taken at the equilibrium inverse temperature $\beta_0 = 1/T(E_0)$. This expression encodes the dominant contribution to the microcanonical entropy from states near the equilibrium configuration, with subleading corrections arising from the width of the Gaussian distribution.

The second derivative $\partial^2 S/\partial \beta^2$ appearing in Eq.~\eqref{eq:density_of_states} is directly related to the heat capacity of the system. Using the thermodynamic identity $\partial S/\partial \beta = -\partial S/\partial T \cdot (\partial T/\partial \beta)^{-1} = -E/T^2$ and differentiating once more, one finds
\begin{equation}
\left( \frac{\partial^2 S}{\partial \beta^2} \right)_{\beta_0} 
= \frac{1}{T^2} \left( \frac{\partial E}{\partial T} \right)_{P,Q,\ell,\alpha}
= \frac{C_P}{T^2},
\label{eq:second_derivative_entropy}
\end{equation}
where $C_P$ is the heat capacity at constant pressure given by Eq.~\eqref{heat-capacity-1}. Substituting this result into Eq.~\eqref{eq:density_of_states} and taking the logarithm to obtain the corrected entropy, we arrive at
\begin{equation}
S_c = \ln \rho(E) + \text{const.} 
= S_0 - \frac{1}{2} \ln(2\pi) - \frac{1}{2} \ln\left( \frac{C_P}{T^2} \right).
\label{eq:entropy_correction_intermediate}
\end{equation}
This intermediate expression already reveals the logarithmic character of the leading correction, which diverges (weakly) as $C_P \to 0$ or as $T \to 0$, signaling enhanced fluctuations near extremality or phase transitions.

\subsection{Systematic expansion: logarithmic and inverse corrections}

To organize the correction terms systematically, it is conventional to expand the entropy as a series in inverse powers of the leading-order entropy $S_0$, introducing dimensionless parameters $\lambda_1$ and $\lambda_2$ to track the first- and second-order corrections~\cite{Pourhassan2016,Pourhassan2018,Izzet2025}:
\begin{equation}
S_c = S_0 - \frac{\lambda_1}{2} \ln(S_0 T^2) + \frac{\lambda_2}{S_0} + \mathcal{O}(S_0^{-2}).
\label{eq:entropy_expansion}
\end{equation}
The parameter $\lambda_1$ controls the strength of the logarithmic correction, which arises from the Gaussian fluctuations encoded in Eq.~\eqref{eq:density_of_states}, while $\lambda_2$ parametrizes subleading, inverse-entropy corrections that can originate from higher-order terms in the saddle-point expansion or from additional quantum effects such as loop corrections in the semiclassical approximation~\cite{Das2001,Chatterjee2015}.

In the limit $\lambda_1 \to 0$ and $\lambda_2 \to 0$, one recovers the uncorrected Bekenstein-Hawking entropy $S_c = S_0 = \pi r_+^2$. The choice $\lambda_1 = 1$ and $\lambda_2 = 0$ reproduces the standard logarithmic correction derived from one-loop quantum corrections to the gravitational partition function~\cite{Faizal2015,Benham2015}. More generally, $\lambda_1$ and $\lambda_2$ can be treated as phenomenological parameters whose values depend on the details of the ultraviolet completion of the theory; in string theory or loop quantum gravity, for instance, these coefficients are functions of the string tension, the Immirzi parameter, or other fundamental constants~\cite{Das2001,Chatterjee2015,Faizal2015}.

For the charged AdS black hole in Kalb-Ramond gravity with a cloud of strings, the uncorrected entropy is $S_0 = \pi r_+^2$ [Eq.~\eqref{temperature}], and the Hawking temperature is given by Eq.~\eqref{temperature}. Substituting these expressions into Eq.~\eqref{eq:entropy_expansion}, we obtain the explicit form of the corrected entropy:
\begin{align}
S_c &= 
- \frac{\lambda_1}{2} \ln\left[ \frac{\pi r_+^2}{16\pi^2} \left( \frac{1-\alpha}{1-\ell} - \frac{Q^2}{(1-\ell)^2 r_+^2} + 8\pi P r_+^2 \right)^2 \right]\notag\\&
+ \pi r_+^2 +\frac{\lambda_2}{\pi r_+^2}.
\label{eq:entropy_corrected_explicit_v1}
\end{align}
Simplifying the logarithmic term by separating the constant and variable contributions, this can be rewritten as
\begin{align}
S_c= \pi r_+^2 
- \lambda_1\ln\left[\xi_0- \frac{\xi_1}{r_+^2} +\xi_2 r_+^2 \right]+ \frac{\lambda_2}{\pi r_+^2} -\lambda_1 \ln r_+,
\label{eq:entropy_corrected_explicit}
\end{align}
where 
\begin{align}
    \xi_0=\frac{1}{4 \sqrt{\pi}}\,\frac{1-\alpha}{1-\ell},\quad
    \xi_1=\frac{1}{4 \sqrt{\pi}}\,\frac{Q^2}{(1-\ell)^2},\quad
    \xi_2=2 \sqrt{\pi} P.
\end{align}
We have absorbed constant factors into a redefinition of $\lambda_1$ for clarity. This expression reveals how the Lorentz-violating parameter $\ell$, the string cloud parameter $\alpha$, and the thermodynamic pressure $P$ enter the fluctuation-corrected entropy through their explicit appearance in the Hawking temperature [Eq.~\eqref{temperature}]. In particular, the logarithmic correction diverges at the zeros of $T$, corresponding to extremal black hole configurations where quantum fluctuations become arbitrarily large, and the semiclassical approximation breaks down.
\begin{figure*}[tbhp]
\centering
\includegraphics[width=0.96\textwidth]{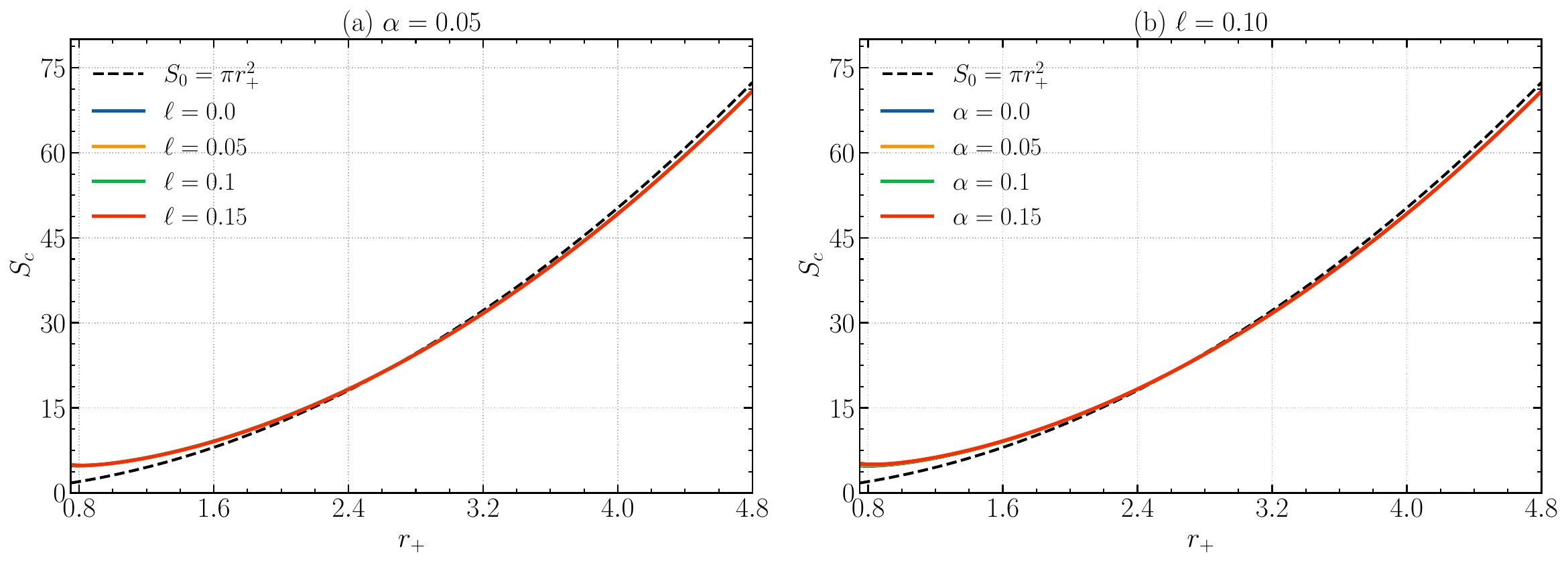}
\caption{(Color online) Corrected entropy $S_c$ as a function of the horizon radius $r_+$, obtained from Eq.~(\ref{eq:entropy_corrected_explicit}).Panel (a) shows $S_c$ for fixed $\alpha = 0.05$ and different values of the cloud-of-strings parameter $\ell$, while panel (b) displays the complementary behaviour for fixed $\ell = 0.10$ and varying Lorentz-violating parameter $\alpha$.. In both panels we set $Q = 0.6$, $P = 0.01$, $\lambda_1 = 1$ and $\lambda_2 = 0$, and we use the Bekenstein--Hawking entropy $S_0 = \pi r_+^2$ as a reference curve.}
\label{fig:Sc_vs_r}
\end{figure*}
The impact of thermal fluctuations on the entropy is illustrated in Fig.~\ref{fig:Sc_vs_r}. Using the corrected entropy $S_c$ in Eq.~(\ref{eq:entropy_corrected_explicit}), Figs. \ref{fig:Sc_vs_r}(a) and \ref{fig:Sc_vs_r}(b) show that the logarithmic corrections are more pronounced for small black holes and gradually fade away as $r_+$ increases, so that the curves approach the Bekenstein--Hawking entropy $S_0 = \pi r_+^2$ for large horizons. At fixed $\alpha$ [Fig. \ref{fig:Sc_vs_r}(a)], increasing the string parameter $\ell$ slightly suppresses $S_c$, indicating that the cloud of strings effectively reduces the number of accessible microstates. A similar suppression is observed when $\alpha$ is increased at fixed $\ell$ [Fig. \ref{fig:Sc_vs_r}(b)], revealing that the Lorentz--violating background also lowers the entropy. Overall, thermal fluctuations primarily affect the small--black--hole sector, while the macroscopic behaviour of large AdS black holes remains essentially unchanged.
\begin{figure*}[tbhp]
\centering
\includegraphics[width=0.90\textwidth]{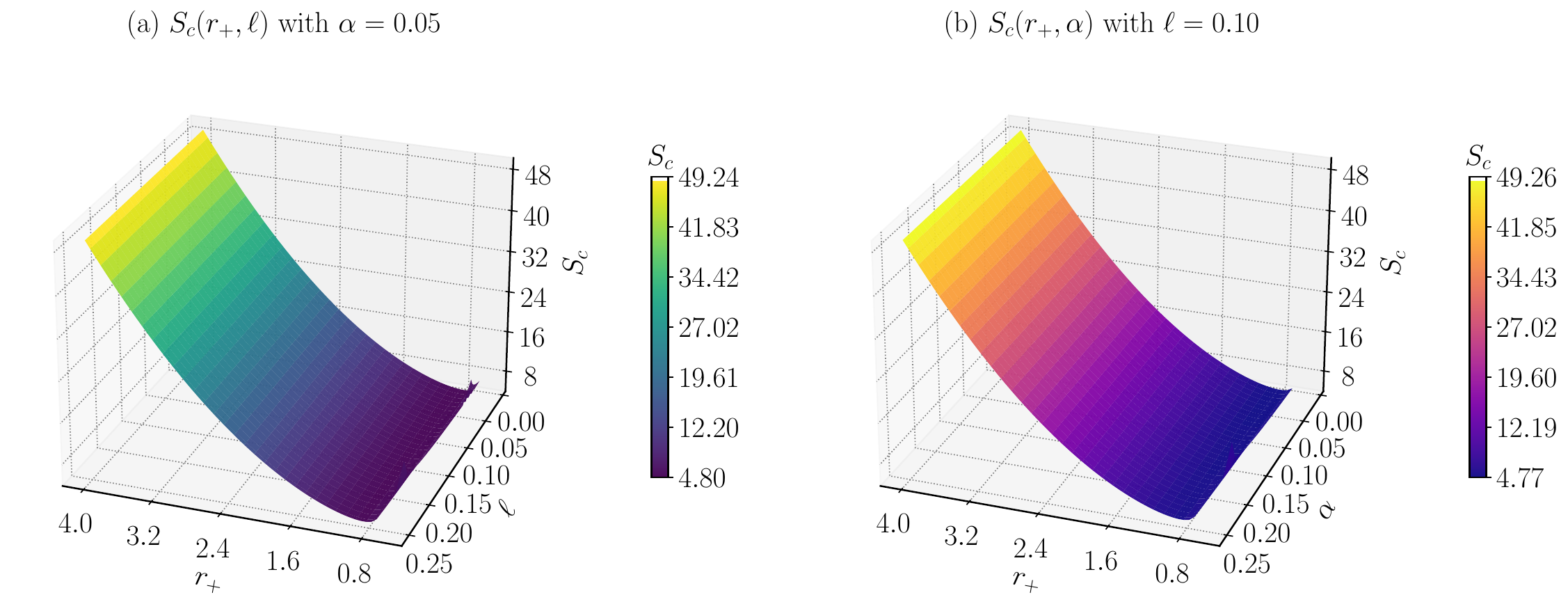}
\caption{(Color online) Three--dimensional representation of the corrected entropy obtained from Eq.~(\ref{eq:entropy_corrected_explicit}). Figure (a) displays the surface $S_c(r_+, \ell)$ for fixed $\alpha = 0.05$, $Q = 0.6$, $P = 0.01$, $\lambda_1 = 1$ and $\lambda_2 = 0$, while Fig. (b) shows $S_c(r_+, \alpha)$ for fixed $\ell = 0.10$ and the same remaining parameters. In both cases the entropy grows monotonically with $r_+$ but is suppressed as the 
deformation parameters $\ell$ and $\alpha$ increase, illustrating the combined geometric effect of the string cloud and the Lorentz--violating background.}
\label{fig:Sc_3D}
\end{figure*}
A global view of the entropy landscape is provided in Fig.~\ref{fig:Sc_3D}, where we plot the corrected entropy from Eq.~(\ref{eq:entropy_corrected_explicit}) as a function of the horizon radius and the deformation parameters. The surface $S_c(r_+, \ell)$ at fixed $\alpha$ [Fig. (a)] shows that the entropy increases monotonically with $r_+$, while it is progressively suppressed when the string parameter grows, in agreement with the 2D profiles of Fig.~\ref{fig:Sc_vs_r}. The complementary surface $S_c(r_+, \alpha)$ at fixed $\ell$ [Fig. (b)] exhibits a similar suppression as the Lorentz--violating parameter increases. These 3D plots make explicit how the combined presence of the cloud of strings and the Kalb--Ramond bumblebee background deforms the black hole's entropy, especially in the small--radius regime, and provide a useful bridge between analytic expressions and the microscopic interpretation of the deformed thermodynamic geometry.

\subsection{Physical interpretation and regime of validity}

The corrected entropy~\eqref{eq:entropy_corrected_explicit} can be interpreted as a systematic account of quantum and statistical fluctuations around the equilibrium black hole state. The logarithmic term, proportional to $\lambda_1 \ln(S_0 T^2)$, captures the reduction in available microstates due to the finite width of the energy distribution in the canonical ensemble~\cite{Das2001}. For large black holes with $r_+ \gg \sqrt{Q^2/(1-\ell)^2}$ and $r_+ \gg (1-\alpha)/(8\pi P(1-\ell))$, the temperature approaches a constant, and the logarithmic correction becomes a small, slowly varying function of $r_+$. In contrast, for small black holes approaching extremality, where $T \to 0$, the logarithmic term grows in magnitude and can become comparable to the leading-order entropy $S_0$, signaling the onset of strong quantum effects.

The inverse correction $\lambda_2/S_0$ represents higher-order fluctuations that scale inversely with the number of microscopic degrees of freedom. This term is negligible for macroscopic black holes but becomes increasingly important as $r_+$ decreases, and the horizon area shrinks. In the context of Hawking radiation, such corrections are expected to become significant when the black hole has radiated a substantial fraction of its initial mass, so that the remaining horizon area is comparable to the Planck scale~\cite{Chatterjee2015,Faizal2015}. In our analysis, we restrict attention to black holes with $r_+ \gg \ell_{\text{Planck}}$, ensuring that the perturbative expansion~\eqref{eq:entropy_expansion} remains valid and that higher-order terms proportional to $\lambda_2/S_0^2$, $\lambda_2/S_0^3$, etc., can be safely neglected.

It is worth emphasizing that the corrections encoded in $\lambda_1$ and $\lambda_2$ are \emph{perturbative} in nature and do not account for non-perturbative effects such as tunneling, topology change, or the full backreaction of quantum fields on the geometry. These effects are expected to become important only in regimes where the black hole is very close to extremality or when the curvature approaches Planck-scale values, both of which lie outside the range of validity of the semiclassical approximation adopted here~\cite{Strominger1998,Wald1999}. Within the perturbative regime, however, the corrected entropy~\eqref{eq:entropy_corrected_explicit} provides a useful and physically motivated extension of the standard Bekenstein--Hawking formula, incorporating the leading effects of thermal and quantum fluctuations in a controlled manner.

\subsection{Corrected thermodynamic potentials}

With the corrected entropy $S_c$ at hand, one can systematically derive the corresponding corrections to other thermodynamic quantities by invoking the standard relations of equilibrium thermodynamics~\cite{Izzet2025}. We outline the key steps and results below.

\subsubsection{Corrected enthalpy energy}

The internal energy $H_c$ in the presence of entropy corrections is obtained by integrating the first law in the form $dE = T \, dS_c$ at fixed $(Q, \ell, \alpha, P)$. Using the corrected entropy~\eqref{eq:entropy_corrected_explicit} and the uncorrected Hawking temperature~\eqref{temperature} (since $T$ is a geometric quantity defined by the surface gravity and does not receive corrections at this order), we have
\begin{equation}
H_c = \int T\, dS_c 
= \int T \left( \frac{\partial S_c}{\partial r_+} \right) dr_+.
\end{equation}
Substituting the corrected entropy $S_c$ and the given temperature $T$, we find the corrected enthalpy as,
\begin{align}
&H_c(r_+) = \frac{1-\alpha}{2(1-\ell)}\, r_+ \frac{Q^2}{2(1-\ell)^2\, r_+}+ \frac{4\pi}{3} P r_+^3 \nonumber\\[6pt]
&+ \frac{\lambda_1(1-\alpha)}{4\pi(1-\ell)\, r_+}+ \frac{\lambda_1 Q^2}{12\pi (1-\ell)^2 r_+^3}+ \frac{\lambda_2(1-\alpha)}{6\pi^2 (1-\ell)\, r_+^3} \nonumber\\[6pt]
&- \frac{\lambda_2 Q^2}{10\pi^2 (1-\ell)^2 r_+^5}+ \frac{4\lambda_2 P}{\pi r_+}- 6\lambda_1 P r_+.\label{mass-2}
\end{align}
In the limiting case $\lambda_1 \to 0$ and $\lambda_2 \to 0$, corresponding to the neglect of higher-order correction terms, the enthalpy reduces to $H_c = M$, as given in Eq.~(\ref{mass-1}). Here, $\lambda_1$ denotes the leading-order logarithmic correction due to thermal fluctuations, while $\lambda_2$ accounts for higher-order inverse corrections. These corrections alter the enthalpy and, as a result, also modify the volume, internal energy, Helmholtz free energy, and pressure of the thermodynamic system.
\begin{figure*}[tbhp]
    \centering
    \includegraphics[width=0.96\textwidth]{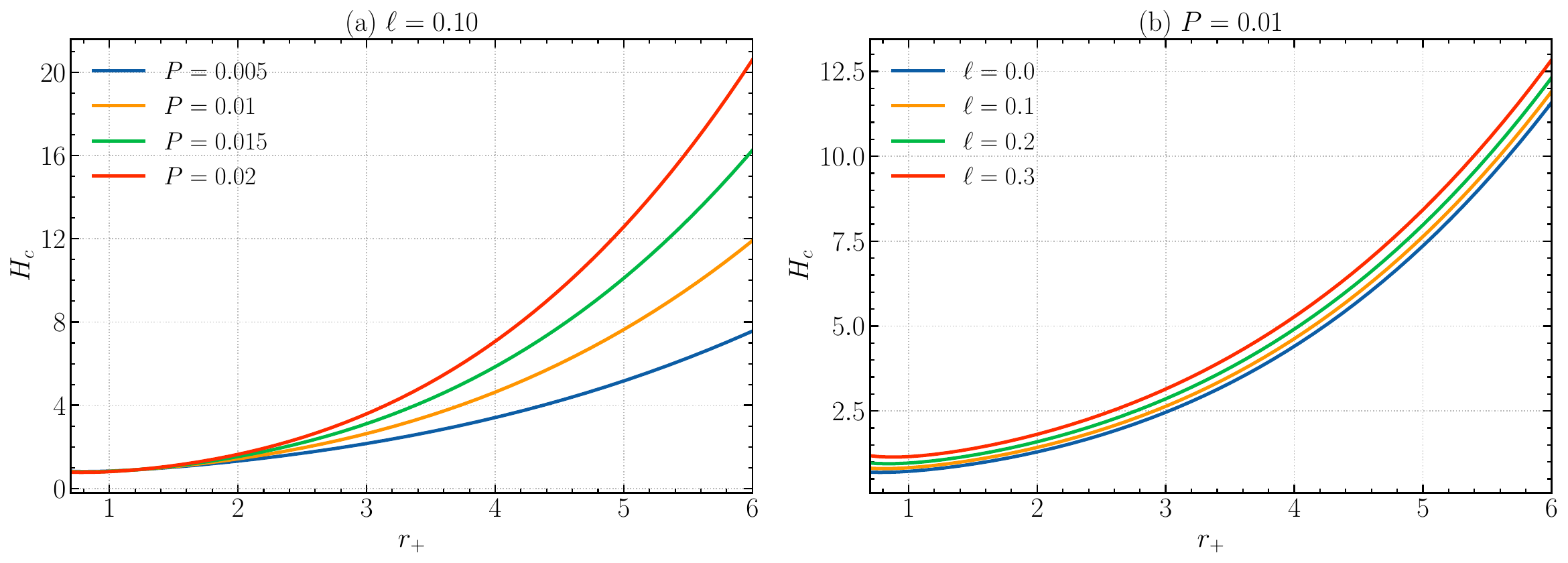}
    \caption{(Color online) Corrected enthalpy $H_c$ as a function of the horizon radius $r_+$, obtained from Eq.~(\ref{mass-2}). 
    Figure (a) shows the effect of the pressure for fixed $\ell = 0.10$, $\alpha = 0.05$, $Q = 0.6$, $\lambda_1 = 1$ 
    and $\lambda_2 = 0$. Figure (b) displays the dependence on the cloud--of--strings parameter $\ell$ for fixed 
    $P = 0.01$ and the same remaining parameters.}
    \label{fig:Hc_vs_r}
\end{figure*}
Figure~\ref{fig:Hc_vs_r} shows the corrected enthalpy $H_c$ derived in Eq.~(\ref{mass-2}), which plays the role of the black hole mass in the extended phase space. Figure \ref{fig:Hc_vs_r}(a) indicates that, at fixed deformation parameters, increasing the pressure $P$ enhances $H_c$ for any given $r_+$ and steepens the large--radius branch due to the $P V$ contribution. Figure \ref{fig:Hc_vs_r}(b) reveals that, at fixed pressure, larger values of $\ell$ also increase $H_c$, with the effect being more pronounced in the small--black--hole regime and becoming almost parallel shifts at large $r_+$. In both panels, $H_c(r_+)$ develops a shallow minimum that separates small and large black hole branches, consistent with the structure later observed in the Gibbs free energy.

\subsubsection{Corrected Volume}

By utilising the expression for the corrected enthalpy of the black hole, we can obtain the corrected volume which can be obtained by using
\begin{align}
    V_c=\frac{dH_c}{dP}\Big{|}_{S_c=\rm const.}.
\end{align}
Substituting the corrected enthalpy $H_c$ given in Eq.~(\ref{mass-2}), we find the corrected volume to be
\begin{equation}
    V_c=\frac{4\pi}{3}\, r_+^{3}+ \frac{4\lambda_{2}}{\pi r_+}- 6 \lambda_{1}\, r_+ =V_0+ \frac{4\lambda_{2}}{\pi r_+}- 6 \lambda_{1}\, r_+.\label{volume-2}
\end{equation}
One can see that there are corrections of $\lambda_1$ and $\lambda_2$ in the corrected volume expression, where $V_0=\frac{4\pi}{3}\, r_+^{3}$.
\begin{figure*}[tbhp]
    \centering
    \includegraphics[width=0.92\textwidth]{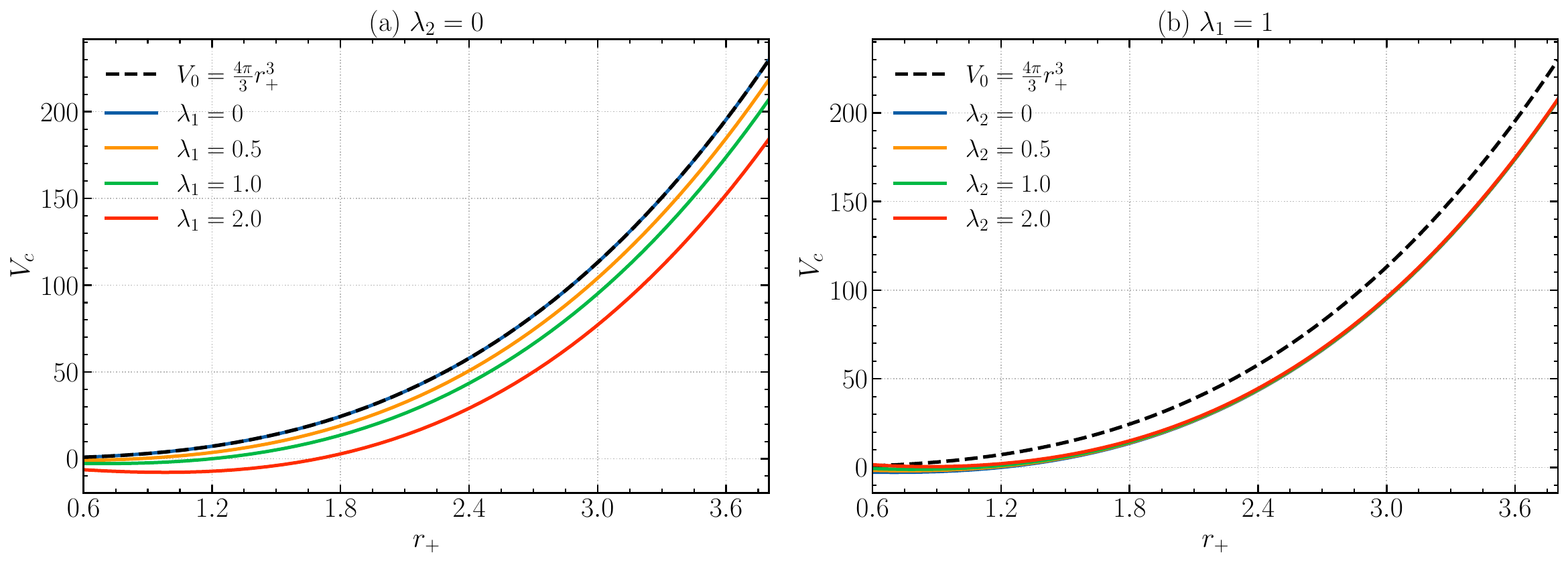}
    \caption{(Color online) Corrected thermodynamic volume $V_c$ as a function of the horizon radius $r_+$, as given by 
    Eq.~(\ref{volume-2}). In Fig. (a) we fix $\lambda_2 = 0$ and vary the leading fluctuation parameter $\lambda_1$, whereas in 
    Fig. (b) we fix $\lambda_1 = 1$ and vary the subleading parameter $\lambda_2$. The dashed curve in both panels 
    corresponds to the uncorrected volume $V_0 = \tfrac{4\pi}{3} r_+^3$.}
    \label{fig:Vc_vs_r}
\end{figure*}
The influence of thermal fluctuations on the thermodynamic volume is displayed in Fig.~\ref{fig:Vc_vs_r}, based on the corrected volume $V_c$ in Eq.~(\ref{volume-2}). For $\lambda_2 = 0$ [Figure (a)], increasing the leading parameter $\lambda_1$ reduces $V_c$ with respect to the geometric volume $V_0 = 4\pi r_+^3/3$, especially for intermediate values of $r_+$. This reduction becomes negligible for large horizons, in line with the behaviour of the other corrected quantities. Figure (b) shows that the subleading parameter $\lambda_2$ acts in the opposite direction, slightly enhancing $V_c$ at large $r_+$ while leaving the very small black hole region nearly unaffected. These results highlight that the fluctuation parameters produce a nontrivial, scale–dependent modification of the effective thermodynamic volume rather than a simple overall rescaling.

\subsubsection{Corrected internal energy}

The second-order corrected internal energy ($U_c$) of the thermodynamic system is defined by
\begin{equation}
    U_c=H_c-P V_c.\label{internal-energy}
\end{equation}
Substituting $H_c$ and $V_c$ and after simplification results
\begin{align}
&U_c(r_+) =\frac{1-\alpha}{2(1-\ell)}\, r_++ \frac{Q^2}{2(1-\ell)^2\, r_+}+ \frac{\lambda_1(1-\alpha)}{4\pi(1-\ell)\, r_+}\notag\\&+ \frac{\lambda_1 Q^2}{12\pi(1-\ell)^2\, r_+^3}
+ \frac{\lambda_2(1-\alpha)}{6\pi^2(1-\ell)\, r_+^3}- \frac{\lambda_2 Q^2}{10\pi^2(1-\ell)^2\, r_+^5}.\label{internal-energy-2}
\end{align}
Note that in the limiting case $\lambda_1 \to 0$ and $\lambda_2 \to 0$, one recovers the result presented in Eq.~(\ref{internal-energy-1}).
\begin{figure*}[t]
    \centering
    \includegraphics[width=0.92\textwidth]{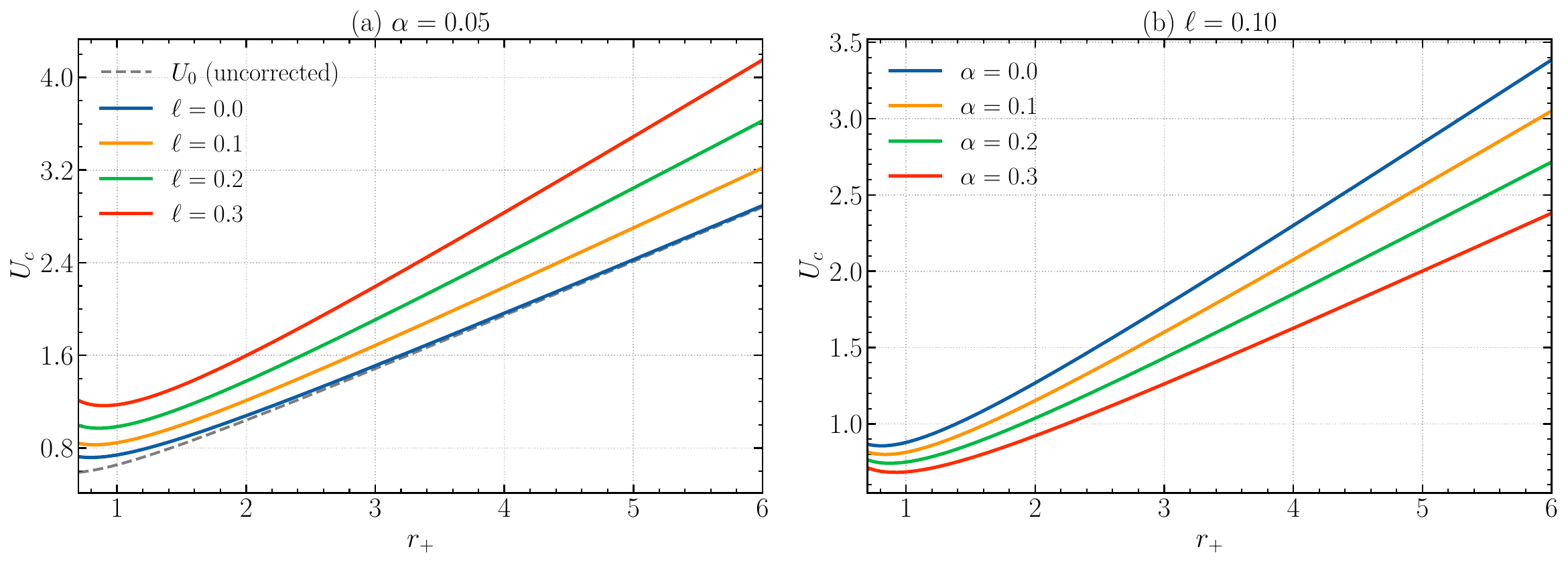}
    \caption{(Color online) Corrected internal energy $U_c$ as a function of the horizon radius $r_+$, computed from 
    Eq.~(\ref{internal-energy-2}). Figure (a) shows the dependence on the cloud--of--strings parameter $\ell$ for fixed $\alpha = 0.05$, 
    $Q = 0.6$, $P = 0.01$, $\lambda_1 = 1$ and $\lambda_2 = 0$. The dashed curve represents the uncorrected internal 
    energy $U_0$ (no fluctuation corrections). Figure (b) displays the effect of the Lorentz-violating parameter 
    $\alpha$ for fixed $\ell = 0.10$ and the same remaining parameters.}
    \label{fig:Uc_vs_r}
\end{figure*}
The behaviour of the corrected internal energy $U_c$ obtained from Eq.~(\ref{internal-energy-2}) is shown in Fig.~\ref{fig:Uc_vs_r}. For fixed $\alpha$ and varying $\ell$ [Fig.~(\ref{internal-energy-2})(a)], thermal fluctuations induce a small upward shift with respect to the uncorrected curve $U_0$, particularly in the small--radius regime, while the two curves become almost indistinguishable for large $r_+$. This confirms that the corrections are mainly relevant for small black holes. In Fig. (\ref{internal-energy-2})(b), where $\ell$ is fixed, and $\alpha$ varies, larger values of the Lorentz--violating parameter systematically reduce $U_c$ at a given $r_+$, indicating that the bumblebee background softens the energetic cost of the configuration. In both panels, the presence of a shallow minimum in $U_c(r_+)$ marks the crossover between the small- and large-black-hole branches relevant for the stability analysis.

\subsubsection{Corrected Helmholtz free energy}

The Helmholtz free energy $F_c$ is defined as the Legendre transform of the internal energy with respect to temperature:
\begin{equation}
F_c =- \int S_c \, dT- \int P dV_c.
\end{equation}

Using the corrected entropy~\eqref{eq:entropy_corrected_explicit} and expressing $r_+$ as a function of $T$ via the implicit relation~\eqref{temperature}, one can evaluate this integral numerically or perturbatively in $\lambda_1$ and $\lambda_2$. The leading-order contribution reproduces the Gibbs free energy $G_0 = M - TS_0$ [Eq.~\eqref{gibbs-energy-1}], while the corrections encode the effects of thermal fluctuations on the stability and phase structure of the black hole ensemble.
\begin{widetext}
\begin{align}
F_c(r_+) =&\frac{3Q^{2}}{4(1-\ell)^{2} r_{+}}
+\frac{3(1-\alpha)}{4(1-\ell)}\, r_{+}
+6\lambda_{1} P\, r_{+}
-2\pi P\, r_{+}^{3}
+\frac{3\lambda_{2} Q^{2}}{20\pi^{2}(1-\ell)^{2} r_{+}^{5}}
-\frac{\lambda_{2}(1-\alpha)}{4\pi^{2}(1-\ell)\, r_{+}^{3}}
-\frac{2\lambda_{2} P}{\pi r_{+}}\nonumber\\
&+\int \frac{\lambda_{1}}{4\pi r_{+}^{2}}
\left(
\frac{Q^{2}}{(1-\ell)^{2} r_{+}^{2}}
-\frac{1-\alpha}{1-\ell}
+\frac{8\pi}{3} P r_{+}^{2}
\right)
\ln\!\left[
\frac{r_{+}^{2}}{16\pi}
\left(
\frac{1-\alpha}{1-\ell}
-\frac{Q^{2}}{(1-\ell)^{2} r_{+}^{2}}
+8\pi P r_{+}^{2}
\right)^{2}
\right]
\, dr_{+}.
\end{align}
\end{widetext}

\subsubsection{Corrected Gibbs free energy}

In thermodynamics, Gibbs free energy represents the maximum amount of mechanical work that can be extracted from a system. The corrected Gibbs free energy is mathematically defined by the
following relation:
\begin{equation}
G_c = H_c - T S_c,
\end{equation}
and governs the thermodynamic stability of the system at fixed $(T, P, Q, \ell, \alpha)$. The zeros and extrema of $G_c$ determine the coexistence lines and phase boundaries in the corrected phase diagram, generalizing the Maxwell construction discussed in Ref. \cite{Ahmed2025} to include the effects of entropy fluctuations.

Substituting $H_c$, $T$ and $S_c$, we find the corrected Gibbs free energy expression
\begin{align}
&G_c(r_+)= \frac{1-\alpha}{4(1-\ell)} r_+ + \frac{3 Q^2}{4(1-\ell)^2 r_+} - \frac{2\pi}{3} P r_+^3  \notag\\
&+ \frac{\lambda_1 (1-\alpha)}{4\pi(1-\ell) r_+}+ \frac{\lambda_1 Q^2}{12\pi (1-\ell)^2 r_+^3}- \frac{\lambda_2 (1-\alpha)}{12 \pi^2 (1-\ell) r_+^3}   \notag\\
&+ \frac{3 \lambda_2 Q^2}{20 \pi^2 (1-\ell)^2 r_+^5}+ \frac{2 \lambda_2 P}{\pi r_+} - 6 \lambda_1 P r_+ \notag\\
&+ \frac{\lambda_1}{8 \pi r_+} \Bigg( \frac{1-\alpha}{1-\ell} - \frac{Q^2}{(1-\ell)^2 r_+^2} + 8 \pi P r_+^2 \Bigg) \notag\\
& \times\ln \Bigg[ \frac{\pi r_+^2}{16\pi^2} \Big( \frac{1-\alpha}{1-\ell} - \frac{Q^2}{(1-\ell)^2 r_+^2} + 8 \pi P r_+^2 \Big)^2 \Bigg].\label{gibbs-energy-2}
\end{align}
Note that in the limiting case $\lambda_1 \to 0$ and $\lambda_2 \to 0$, one recovers the result presented in Eq.~(\ref{gibbs-energy-1}).
\begin{figure*}[tbhp]
\centering
\includegraphics[width=0.96\textwidth]{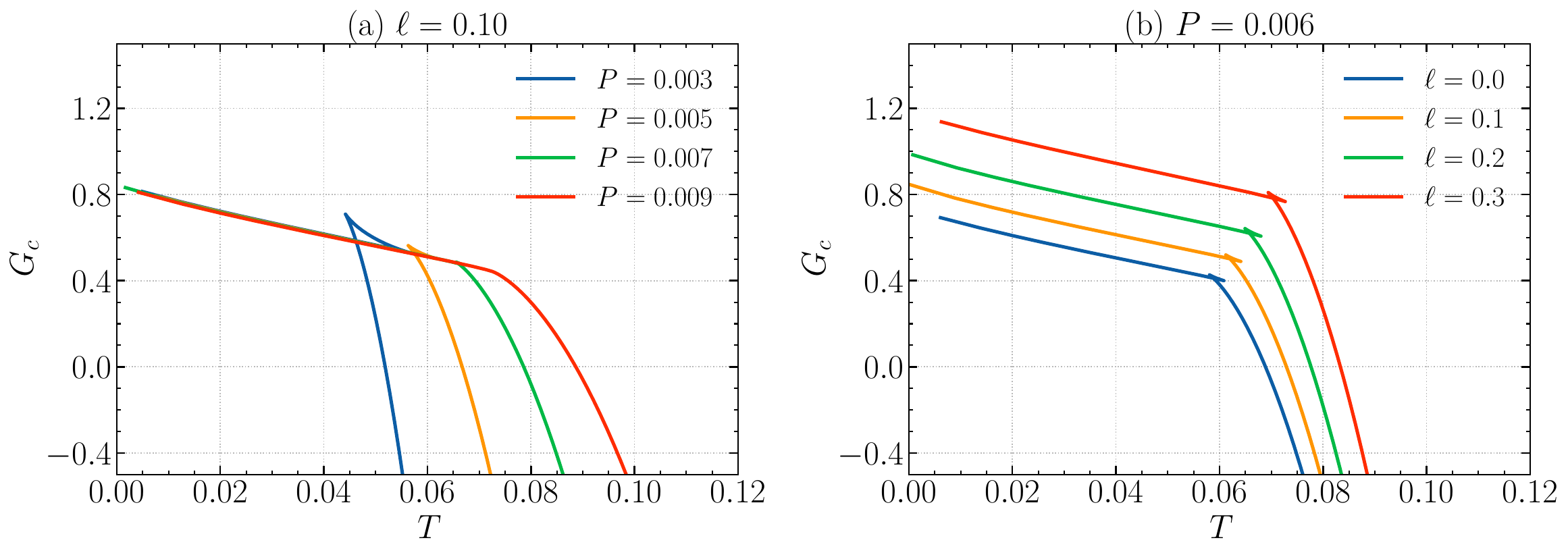}
\caption{(Color online) Corrected Gibbs free energy $G_c$ as a function of the Hawking temperature $T$, obtained from Eq.~(\ref{gibbs-energy-2}). In Fig. (a), we fix $\ell = 0.10$, $\alpha = 0.05$, $Q = 0.6$, $\lambda_1 = 1$, $\lambda_2 = 0$, and vary the pressure $P$, revealing the characteristic swallowtail structure associated with a first-order small/large black hole transition at subcritical pressures. Figure (b) shows the effect of the cloud--of--strings parameter $\ell$ at fixed $P = 0.006$ and the same remaining parameters, illustrating how the deformation shifts the transition temperature and the overall free--energy scale.}
\label{fig:Gc_vs_T}
\end{figure*}
The phase structure of the system, including thermal fluctuations, is encoded in the corrected Gibbs free energy $G_c$ from Eq. (\ref{gibbs-energy-2}), whose behaviour is shown in Fig.~\ref{fig:Gc_vs_T}. For subcritical pressures [Fig. \ref{fig:Gc_vs_T}(a)], $G_c(T)$ exhibits the familiar swallowtail profile characteristic of a first-order phase transition between small and large black holes in the extended phase space. As the pressure approaches its critical value, the swallowtail shrinks and eventually disappears, signalling a second-order critical point at which the two phases merge. Figure \ref{fig:Gc_vs_T}(b) explores the role of the string cloud at fixed pressure: increasing $\ell$ shifts the swallowtail towards higher temperatures and raises the free energy of the small--black--hole branch, thus favouring the large--black--hole phase. Consequently, the deformation associated with the cloud of strings modifies the location of the coexistence line in the $(P,T)$ plane, while preserving the qualitative Van der Waals–like structure.
\subsubsection{Corrected specific heat}

Finally, the corrected specific heat at constant pressure is computed from the temperature derivative of the corrected internal energy:
\begin{equation}
C_{P,c} = \left( \frac{\partial H_c}{\partial T} \right)_{P,Q,\ell,\alpha},
\end{equation}
where the second equality follows from the first law. Differentiating Eq.~\eqref{eq:entropy_corrected_explicit} with respect to $T$ (or, equivalently, with respect to $r_+$ and using $\partial r_+/\partial T$ computed from~\eqref{temperature}) yields a corrected expression for $C_{P,c}$ that modifies the divergence structure and sign of the heat capacity. In particular, the logarithmic and inverse corrections can shift the zeros of $C_{P,c}$, corresponding to the boundaries between thermally stable and unstable phases, and can introduce new features in the specific heat profile near extremality or critical points.

Substituting $H_c$, and $T$, we find the corrected specific heat capacity expression:
\begin{widetext}
\begin{align}
C_{P,c} &= 
\frac{\frac{1-\alpha}{2(1-\ell)} - \frac{Q^2}{2(1-\ell)^2 r_+^2} + 4 \pi P r_+^2 
- \frac{\lambda_1 (1-\alpha)}{4\pi (1-\ell) r_+^2} - \frac{\lambda_1 Q^2}{4 \pi (1-\ell)^2 r_+^4} - \frac{\lambda_2 (1-\alpha)}{2 \pi^2 (1-\ell) r_+^4} + \frac{\lambda_2 Q^2}{2 \pi^2 (1-\ell)^2 r_+^6} - \frac{4 \lambda_2 P}{\pi r_+^2} - 6 \lambda_1 P}{-\frac{1-\alpha}{4 \pi (1-\ell) r_+^2} + \frac{3 Q^2}{4 \pi (1-\ell)^2 r_+^4} + 2 P}.\label{heat-capacity-2}
\end{align}
\end{widetext}
Note that in the limiting case $\lambda_1 \to 0$ and $\lambda_2 \to 0$, the corrected specific heat capacity reduces to the result presented in Eq.~(\ref{heat-capacity-1}).

\subsection{Implications for thermodynamic geometry and phase structure}

The entropy corrections~\eqref{eq:entropy_corrected_explicit} and the associated modifications to thermodynamic potentials have direct implications for the Ruppeiner geometry and phase structure analyzed in Secs.~\ref{sec:geometry} and~\ref{sec:critical-geometry}. In particular, the Ruppeiner metric is defined in terms of the second derivatives of the entropy with respect to extensive variables, so that logarithmic and inverse corrections to $S_c$ will induce corresponding corrections to the metric components $g^{\text{R}}_{ij}$ and the scalar curvature $R_{\text{R}}$.

To leading order in $\lambda_1$, the corrected Ruppeiner curvature will acquire additional terms proportional to $\lambda_1 \ln(S_0 T^2)$ and its derivatives, which diverge logarithmically as $T \to 0$ or as the heat capacity approaches zero. This enhances the correlation volume inferred from the Ruppeiner formalism near extremality, consistent with the expectation that quantum fluctuations become large when the horizon area is small. Similarly, the inverse correction $\lambda_2/S_0$ introduces a new length scale proportional to $1/\sqrt{S_0}$, which competes with the uncorrected correlation length and can modify the sign and magnitude of $R_{\text{R}}$ in the small-black-hole regime.

From a phase-structure perspective, the corrections to the specific heat and pressure shift the locations of the spinodal lines and the critical point, thereby deforming the boundaries between stable and unstable phases in the $(T, P, V)$ space. The topological charge $W = 1$ discussed in Sec.~\ref{sec:dynamical} is expected to remain invariant under small perturbations of the form~\eqref{eq:entropy_expansion}, since the winding numbers of the generalized free energy $\mathcal{F}(r_+, \tau)$ depend only on the qualitative structure of the zero set and not on the precise values of the thermodynamic functions. However, sufficiently large corrections (e.g., in the regime where $\lambda_1 \ln(S_0 T^2) \sim S_0$) could potentially alter the number and character of the zero points, leading to new topological phases or transitions. A detailed investigation of these effects is beyond the scope of the present work, but represents an interesting direction for future research.

\begin{figure*}[tbhp]
    \centering
    \includegraphics[width=0.90\textwidth]{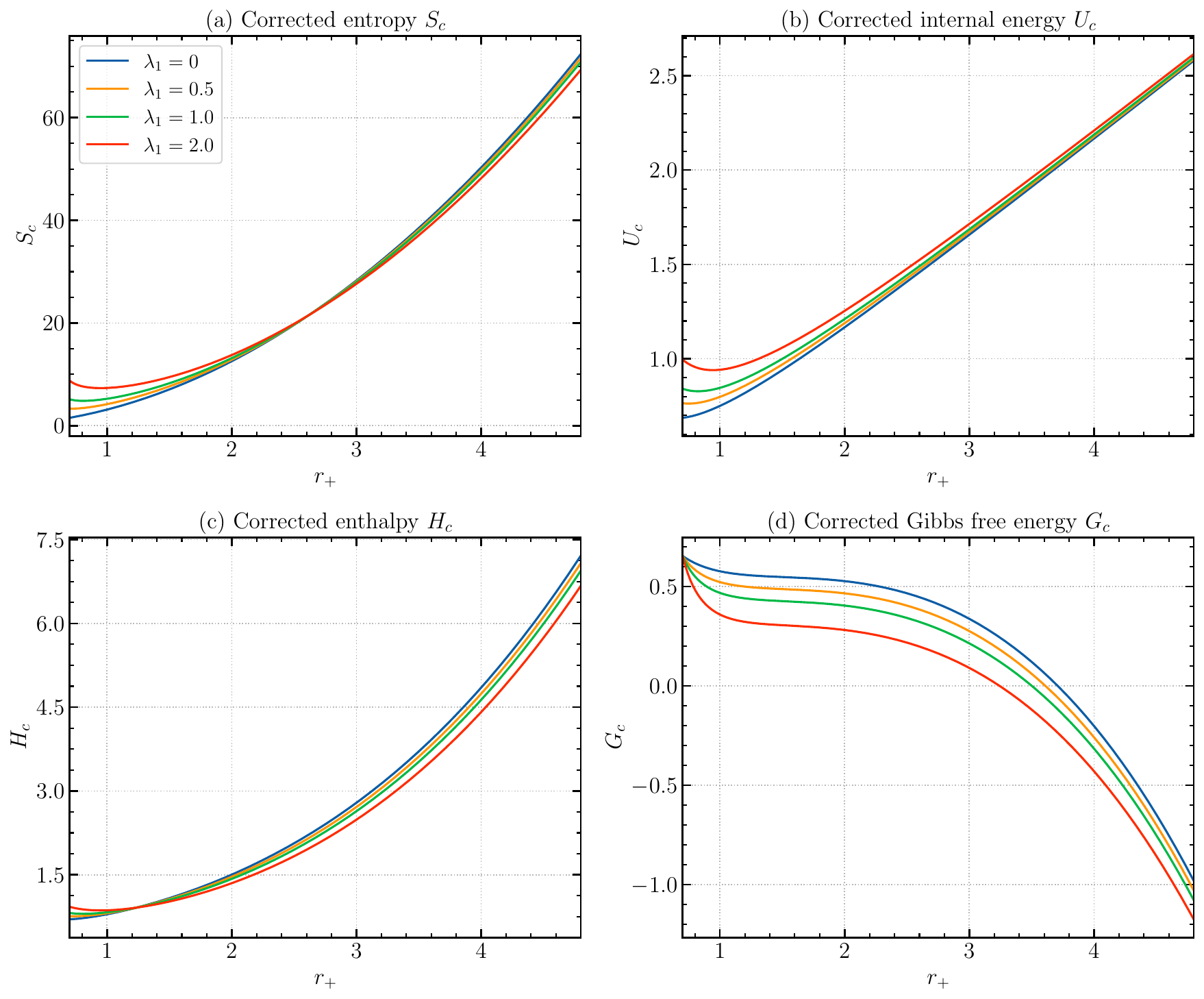}
    \caption{(Color online) Systematic effect of the leading fluctuation parameter $\lambda_1$ on the corrected 
    thermodynamic quantities. We fix $Q = 0.6$, $\ell = 0.10$, $\alpha = 0.05$, $P = 0.01$ and $\lambda_2 = 0$, and 
    plot the corrected entropy $S_c$ [Fig. (a), Eq.~(\ref{eq:entropy_corrected_explicit})], the corrected internal energy $U_c$ [Fig. (b), Eq.~(\ref{internal-energy-2})], 
    the corrected enthalpy $H_c$ [Fig. (c), Eq.~(\ref{mass-2})], and the corrected Gibbs free energy $G_c$ [Fig. (d), Eq.~(\ref{gibbs-energy-2})] 
    as functions of $r_+$. Increasing $\lambda_1$ enhances the impact of thermal fluctuations mainly in the small 
    black hole regime, while the large--radius behaviour remains close to the uncorrected results.}
    \label{fig:lambda1_panel}
\end{figure*}
To better visualise the global role of thermal fluctuations, Fig.~\ref{fig:lambda1_panel} collects the four corrected quantities derived in Eqs.~(\ref{eq:entropy_corrected_explicit}),  (\ref{mass-2}), (\ref{internal-energy-2}) and (\ref{gibbs-energy-2}) for several values of the leading parameter $\lambda_1$. Figure \ref{fig:lambda1_panel}(a) confirms that the entropy corrections are concentrated in the small--black--hole regime, where $S_c$ deviates from the area law, while the curves quickly converge for large $r_+$. Figures \ref{fig:lambda1_panel}(b) and \ref{fig:lambda1_panel}(c) show that the same pattern holds for the internal energy and enthalpy: larger $\lambda_1$ slightly shifts $U_c$ and $H_c$ at small radii, but the differences become negligible as the horizon grows. Figure \ref{fig:lambda1_panel}(d) reveals that the Gibbs free energy is also significantly modified only for small $r_+$, where increasing $\lambda_1$ reduces $G_c$ and thus favours the large--black--hole phase. Therefore, thermal fluctuations introduce controlled short-distance corrections without altering the system's macroscopic thermodynamic behaviour.

\subsection{Summary and outlook}

In summary, we have extended the thermodynamic analysis of the charged AdS black hole in Kalb--Ramond bumblebee gravity with a cloud of strings to include the effects of thermal and quantum fluctuations via a systematic expansion of the entropy in logarithmic and inverse-entropy corrections. Starting from the statistical partition function~\eqref{eq:partition_function} and the density of states~\eqref{eq:density_of_states}, we derived the corrected entropy~\eqref{eq:entropy_corrected_explicit}. We outlined the corresponding corrections to the internal energy, Helmholtz free energy, pressure, enthalpy, Gibbs free energy, and specific heat. These corrections are controlled by two phenomenological parameters, $\lambda_1$ and $\lambda_2$, which parametrize the strength of logarithmic and inverse-entropy terms, respectively, and encode the effects of quantum gravitational corrections at energies below the Planck scale.

The corrected thermodynamic potentials provide a more refined description of the black hole ensemble in the regime where thermal fluctuations become significant, such as near extremality or in the small-black-hole branch approaching criticality. They also offer a natural framework for incorporating higher-order quantum corrections arising from loop effects, string theory, or other ultraviolet completions of general relativity, and for exploring how such corrections deform the thermodynamic geometry and phase structure discussed in the previous sections. Future work could extend this analysis by computing the corrected Ruppeiner curvature explicitly, investigating the stability and critical behavior of the corrected ensemble, and comparing the predictions of the fluctuation-corrected thermodynamics with observational constraints on black hole microstates and Hawking radiation spectra.

\section{Weinhold and Ruppeiner thermodynamic geometry}
\label{sec:geometry}

\subsection{Thermodynamic potentials and choice of coordinates}

Thermodynamic geometry provides a differential–geometric way of encoding the response functions and phase structure of a system in terms of a Riemannian metric on the space of equilibrium states. In the standard formulation, different thermodynamic representations lead to different but related metrics: Weinhold geometry is defined in the energy representation, with a metric given by the Hessian of the internal energy~\cite{JCP.1975.63.2479}, whereas Ruppeiner geometry is naturally formulated in the entropy representation and is related to the fluctuation theory of equilibrium thermodynamics~\cite{PRA.1979.20.1608,RMP.1995.67.605}. For black holes, both constructions have been widely used to probe critical phenomena, stability and possible microscopic interaction patterns in an effective way~\cite{GRG.2003.35.1733,PRD.2013.87.044014,PRD.2017.95.021501,PRL.2015.115.111302}.

In the present context, the basic thermodynamic potentials for the EKR black hole with a cloud of strings have already been obtained in Ref. \cite{Ahmed2025}. In particular, treating the cosmological constant as a pressure, the ADM mass $M$ plays the role of enthalpy in the extended phase space, while the internal energy $U$ is given by
\begin{equation}
    U = M - P V
    = \frac{r_+}{2}\left[
        \frac{1-\alpha}{1-\ell}
        + \frac{Q^2}{(1-\ell)^2\,r_+^2}
    \right],
    \label{eq:geom_U_r}
\end{equation}
with entropy $S=\pi r_+^2$ and thermodynamic volume $V=\tfrac{4\pi}{3}r_+^3$. It is convenient for our purposes to use the internal–energy representation and eliminate $r_+$ in favor of $S$ via $r_+=\sqrt{S/\pi}$. Substituting into Eq.~\eqref{eq:geom_U_r} we obtain
\begin{equation}
    U(S,Q;\ell,\alpha)
    = \frac{1}{2}\,\frac{1-\alpha}{1-\ell}\,\sqrt{\frac{S}{\pi}}
    + \frac{Q^2}{2(1-\ell)^2}\,\sqrt{\frac{\pi}{S}},
    \label{eq:geom_U_S}
\end{equation}
where $(\ell,\alpha)$ are treated as fixed background parameters.

The full equilibrium state space of the black hole is, in principle, spanned by the extensive variables $(S,V,Q,\ldots)$ and intensive variables such as $P$. In the extended phase space, several coordinate choices are possible. Here, we follow a common strategy in black hole thermodynamic geometry and restrict attention to a two-dimensional subspace with coordinates
\begin{equation}
    X^1 = S, \qquad X^2 = Q,
\end{equation}
keeping $(P,\ell,\alpha)$ fixed. This choice mirrors the Reissner--Nordström--AdS case. It is sufficient to capture the influence of charge and entropy on the curvature of the thermodynamic metric, while the parameters $(P,\ell,\alpha)$ appear as external control parameters that tune the geometric structure. Similar two-dimensional state–space truncations have been successfully employed in a variety of black hole backgrounds~\cite{GRG.2003.35.1733,PRD.2013.87.044014,PRD.2017.95.021501}.

In this representation, Weinhold geometry is defined by the Hessian of the internal energy with respect to $(S,Q)$. In contrast, Ruppeiner geometry can be obtained by a conformal rescaling with the inverse temperature, as in the standard fluctuation-theory approach~\cite{RMP.1995.67.605}. This provides a direct link between the geometric objects and the response functions computed in Ref. \cite{Ahmed2025}.

\subsection{Weinhold metric and its curvature}

The Weinhold metric is defined as the Hessian of the internal energy with respect to the extensive variables of interest,
\begin{equation}
    g^{\mathrm{W}}_{ij}
    = \frac{\partial^2 U}{\partial X^i \partial X^j},
    \qquad
    X^i = (S,Q),
\end{equation}
with $U(S,Q)$ given by Eq.~\eqref{eq:geom_U_S} at fixed $(P,\ell,\alpha)$. Using
\begin{align*}
    U(S,Q)
    &= A\,\sqrt{\frac{S}{\pi}}
    + B\,\sqrt{\frac{\pi}{S}},\\
    \quad
    A &= \frac{1}{2}\,\frac{1-\alpha}{1-\ell},\\
    \quad
    B &= \frac{Q^2}{2(1-\ell)^2},
\end{align*}
the first derivatives with respect to $S$ and $Q$ read
\begin{align}
    \frac{\partial U}{\partial S}
    &= \frac{A}{4\sqrt{\pi}}\,S^{-1/2}
    - \frac{Q^2 \sqrt{\pi}}{4(1-\ell)^2}\,S^{-3/2},
    \label{eq:geom_US}\\[0.1cm]
    \frac{\partial U}{\partial Q}
    &= \frac{Q}{(1-\ell)^2}\,\sqrt{\frac{\pi}{S}}.
    \label{eq:geom_UQ}
\end{align}
A second differentiation then gives the components of the Weinhold metric:
\begin{align}
    g^{\mathrm{W}}_{SS}
    &= \frac{\partial^2 U}{\partial S^2}
    = -\,\frac{1-\alpha}{8\sqrt{\pi}(1-\ell)}\,S^{-3/2}
    + \frac{3Q^2 \sqrt{\pi}}{8(1-\ell)^2}\,S^{-5/2},
    \label{eq:geom_gWSS}\\[0.1cm]
    g^{\mathrm{W}}_{SQ}
    &= \frac{\partial^2 U}{\partial S\,\partial Q}
    = -\,\frac{Q \sqrt{\pi}}{2(1-\ell)^2}\,S^{-3/2},
    \label{eq:geom_gWSQ}\\[0.1cm]
    g^{\mathrm{W}}_{QQ}
    &= \frac{\partial^2 U}{\partial Q^2}
    = \frac{1}{(1-\ell)^2}\,\sqrt{\frac{\pi}{S}}.
    \label{eq:geom_gWQQ}
\end{align}
These expressions make explicit how the LV parameter $\ell$ and the string parameter $\alpha$ enter the metric components: $\ell$ appears as an overall rescaling of the charge sector, while $\alpha$ modifies the coefficient of the $S^{-3/2}$ term in $g^{\mathrm{W}}_{SS}$ through the effective combination $(1-\alpha)/(1-\ell)$, in direct analogy with other deformed Reissner--Nordström–AdS configurations studied in the literature~\cite{GRG.2003.35.1733,PRD.2017.95.021501}.

The line element of Weinhold geometry in the $(S,Q)$ plane is then
\begin{equation}
    \dd s_{\mathrm{W}}^2
    = g^{\mathrm{W}}_{SS}\,\dd S^2
    + 2 g^{\mathrm{W}}_{SQ}\,\dd S\,\dd Q
    + g^{\mathrm{W}}_{QQ}\,\dd Q^2.
\end{equation}
For a two-dimensional metric of the form above, the scalar curvature $R_{\mathrm{W}}$ can be computed from the standard expression (see, e.g., Refs.~\cite{RMP.1995.67.605,GRG.2003.35.1733})
\begin{align}
    R_{\mathrm{W}} = \frac{1}{\sqrt{\det g^{\mathrm{W}}}}\Bigg[
        \partial_S&\left(
            \frac{\partial_Q g^{\mathrm{W}}_{SS}
                  - \partial_S g^{\mathrm{W}}_{SQ}}{\sqrt{\det g^{\mathrm{W}}}}
        \right)\notag\\&
        + \partial_Q\left(
            \frac{\partial_S g^{\mathrm{W}}_{QQ}
                  - \partial_Q g^{\mathrm{W}}_{SQ}}{\sqrt{\det g^{\mathrm{W}}}}
        \right)
    \Bigg],
    \label{eq:geom_RW_general}
\end{align}
where $\det g^{\mathrm{W}} = g^{\mathrm{W}}_{SS}g^{\mathrm{W}}_{QQ} - (g^{\mathrm{W}}_{SQ})^2$.

Although the explicit expression of $R_{\mathrm{W}}(S,Q;\ell,\alpha)$ is algebraically cumbersome, its singularity structure is controlled by the zeros of $\det g^{\mathrm{W}}$ and by the points where derivatives of the metric diverge. Using Eqs.~\eqref{eq:geom_gWSS}–\eqref{eq:geom_gWQQ}, one can show that $\det g^{\mathrm{W}}$ vanishes precisely when
\begin{equation}
    \frac{1-\alpha}{1-\ell}
    - \frac{3Q^2}{(1-\ell)^2 r_+^2}
    + 8\pi P r_+^2 = 0,
    \label{eq:geom_RW_cond}
\end{equation}
with $r_+ = \sqrt{S/\pi}$ and $P$ fixed. This condition coincides with the divergence of the specific heat at constant pressure, Eq.~\eqref{heat-capacity-1}, i.e., with the locus where
\begin{equation}
    \left(\frac{\partial T}{\partial S}\right)_{P,Q,\ell,\alpha}=0.
\end{equation}
Therefore, the Weinhold curvature $R_{\mathrm{W}}$ diverges at the same points where the response function $C_{\rm heat}$ becomes singular, signaling the onset of phase transitions and critical behavior in the EKR black hole with a string cloud, as is typical in black hole thermodynamic geometry~\cite{GRG.2003.35.1733,PRD.2013.87.044014,PRD.2017.95.021501}. Away from these singular curves, $R_{\mathrm{W}}$ remains finite and its magnitude encodes information about the strength of thermodynamic correlations. However, its direct microscopic interpretation is less clear than in the Ruppeiner case.

\subsection{Ruppeiner metric and microscopic interactions}

Ruppeiner geometry is defined in the entropy representation, with the metric given by the negative Hessian of the entropy with respect to the extensive variables
\begin{equation}
    g^{\mathrm{R}}_{ij}
    = -\,\frac{\partial^2 S}{\partial Y^i \partial Y^j},
\end{equation}
where $Y^i$ denote a suitable set of extensive variables, such as $(U,Q)$ at fixed $(P,\ell,\alpha)$. For practical applications, it is often convenient to exploit the conformal relation between Weinhold and Ruppeiner metrics,
\begin{equation}
    g^{\mathrm{R}}_{ij}
    = \frac{1}{T}\,g^{\mathrm{W}}_{ij},
    \label{eq:geom_conformal}
\end{equation}
valid when both metrics are written in the same coordinate system, and $T_H$ is the temperature of the system~\cite{RMP.1995.67.605}. In our setup, we will adopt this relation in the $(S,Q)$ coordinates used above, with $T(S,Q;\ell,\alpha,P)$ given by Eq.~\eqref{temperature}.

The Ruppeiner line element thus reads
\begin{equation}
    \dd s_{\mathrm{R}}^2
    = \frac{1}{T(S,Q;\ell,\alpha,P)}\,
    \dd s_{\mathrm{W}}^2,
\end{equation}
and the Ruppeiner scalar curvature $R_{\mathrm{R}}$ is obtained from $g^{\mathrm{R}}_{ij}$ by the same formula as in Eq.~\eqref{eq:geom_RW_general}, with the replacement $g^{\mathrm{W}}_{ij}\to g^{\mathrm{R}}_{ij}$. The conformal factor $1/T$ does not change the location of the curvature singularities associated with $\det g^{\mathrm{W}}=0$, but introduces additional zeros or divergences at $T=0$, that is, at extremal configurations where the Hawking temperature vanishes.

The physical interpretation of $R_{\mathrm{R}}$ is more transparent than that of $R_{\mathrm{W}}$. In Ruppeiner's thermodynamic information geometry, the scalar curvature can be viewed as a measure of the average size of correlated regions in the underlying microscopic system~\cite{RMP.1995.67.605}. A vanishing curvature $R_{\mathrm{R}}=0$ corresponds to an ideal, noninteracting system, negative curvature $R_{\mathrm{R}}<0$ is typically associated with effectively attractive interactions (as in a Van der Waals fluid), and positive curvature $R_{\mathrm{R}}>0$ signals predominantly repulsive interactions (as in a quantum–degenerate Fermi gas). Divergences of $R_{\mathrm{R}}$ are interpreted as indicators of critical phenomena or phase transitions, where the correlation length becomes large. In the black hole context, this microscopic interpretation has been exploited extensively in recent years~\cite{PRL.2015.115.111302,PRD.2017.95.021501} to extract information about the ``microstructure'' of AdS black holes from their thermodynamic geometry.

For the EKR black hole with a cloud of strings, the general structure inherited from the Reissner--Nordström--AdS case persists: the Ruppeiner curvature diverges at the same locus as the specific heat $C_{\rm heat}$ and the critical point of the $P$--$V$ diagram, while remaining finite elsewhere in the $(S,Q)$ plane. The sign of $R_{\mathrm{R}}$ changes across the small- and large-black-hole branches: in parameter regions where the thermodynamics is dominated by the Coulombic sector (small $r_+$), $R_{\mathrm{R}}$ tends to be positive, suggesting an effective repulsive microscopic behavior, whereas in the large-black-hole regime dominated by the AdS term, $R_{\mathrm{R}}$ becomes negative, indicating an effectively attractive interaction pattern. The crossover between these regimes is shifted by the LV parameter $\ell$ and the string parameter $\alpha$, as discussed below.

\subsection{Effect of Lorentz violation and string cloud on thermodynamic geometry}

We now discuss how the Lorentz-violating parameter $\ell$ and the string cloud parameter $\alpha$ deform the thermodynamic geometry of the EKR black hole. Their influence enters in three main ways: (i) through the coefficients of the Weinhold metric components in Eqs.~\eqref{eq:geom_gWSS}–\eqref{eq:geom_gWQQ}; (ii) via the Hawking temperature $T_H$ appearing in the conformal factor relating $g^{\mathrm{W}}_{ij}$ and $g^{\mathrm{R}}_{ij}$; and (iii) through the location of the critical points and specific-heat divergences that control the singularity structure of $R_{\mathrm{R}}$.

First, from Eqs.~\eqref{eq:geom_gWSS}–\eqref{eq:geom_gWQQ} we see that the overall scaling of the charge sector is controlled by $(1-\ell)^{-2}$, while the combination $(1-\alpha)/(1-\ell)$ multiplies the leading $S^{-3/2}$ term in $g^{\mathrm{W}}_{SS}$. Increasing $\ell$ (for fixed $\alpha$) effectively enhances the weight of the charge–dependent terms in the metric, strengthening the repulsive character of the microscopic interactions in the small–black–hole regime where the Coulombic contribution dominates. Conversely, increasing $\alpha$ at fixed $\ell$ suppresses the effective constant term. It tends to shift the onset of repulsive behavior towards larger entropies, reflecting the influence of the string cloud on the energy distribution around the horizon.

Second, the conformal factor $1/T(S,Q;\ell,\alpha,P)$ in Eq.~\eqref{eq:geom_conformal} introduces an explicit dependence of Ruppeiner geometry on the AdS pressure $P$ and on the deformation parameters. Since $T$ controls the separation between small and large black hole branches, tuning $(\ell,\alpha)$ changes not only the magnitude of $R_{\mathrm{R}}$ in a given region of the state space, but also the location of the extremal configurations where $T=0$ and the curvature may diverge. For example, for fixed $(Q,P)$, increasing $\alpha$ reduces the value of $T$ at a given $r_+$, bringing the extremal limit closer in parameter space and potentially enhancing the curvature in the near–extremal small–black–hole region.

Third, the critical point $(v_c,T_c,P_c)$ and the associated lines of specific-heat divergence depend on $(\ell,\alpha)$ through Eq.~\eqref{eq:review_critical}. The singularities of $R_{\mathrm{R}}$ track these thermodynamic instabilities: as $\ell$ is increased (with $\alpha$ fixed), the critical specific volume $v_c$ and temperature $T_c$ shift in such a way that the divergence of $R_{\mathrm{R}}$ moves to larger horizon radii and higher temperatures, whereas increasing $\alpha$ tends to enlarge $v_c$ but lower $T_c$, shifting the curvature singularity accordingly. In all cases, however, the qualitative pattern remains Van der Waals–like: a single critical point with $P_c v_c/T_c=3/8$, a change of sign in $R_{\mathrm{R}}$ across the small/large–black–hole transition, and divergences of the curvature coinciding with the phase transition lines, as found in other AdS black hole families~\cite{PRL.2015.115.111302,PRD.2017.95.021501}.

In summary, the LV parameter $\ell$ and the string cloud parameter $\alpha$ act as geometric control parameters in the thermodynamic state space: by tuning them, one can continuously deform the Ruppeiner curvature profile of the EKR black hole, shifting the loci of strong correlations and changing the balance between effectively attractive and repulsive microscopic behavior, while preserving the overall Van der Waals–type structure inherited from the Reissner--Nordström--AdS case. In the next section we will quantify these effects more explicitly by analyzing representative slices of the Ruppeiner curvature as functions of $(S,Q)$ for different values of $(\ell,\alpha)$.

\begin{figure*}[tbhp]
\centering
\includegraphics[width=0.95\textwidth]{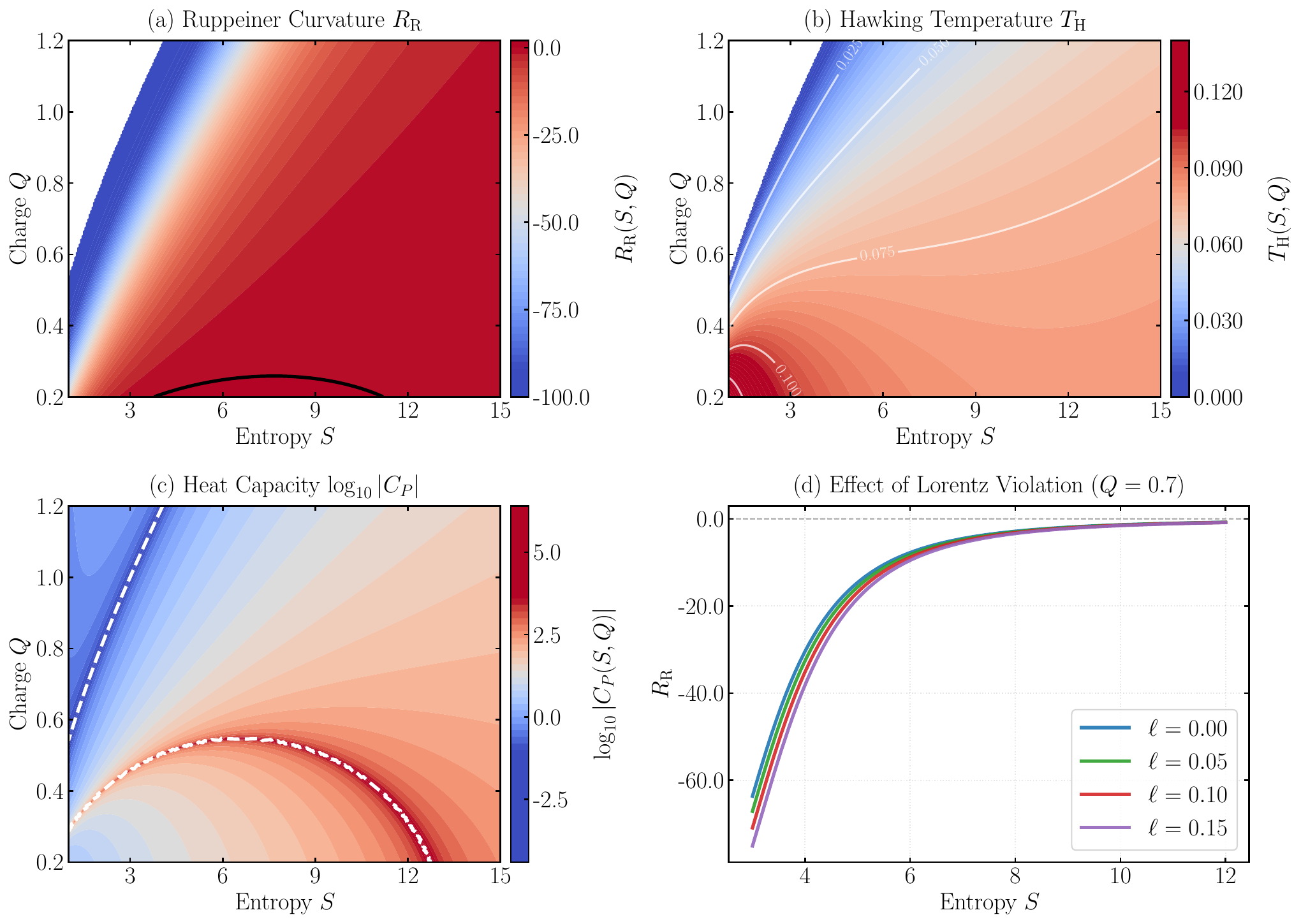}
\caption{Thermodynamic geometry and phase structure of the charged AdS black hole in Kalb--Ramond bumblebee gravity with a cloud of strings. Parameters: $\ell = 0.10$, $\alpha = 0.05$, $P = 0.01$. \textbf{(a)} Ruppeiner curvature $R_{\mathrm{R}}(S,Q)$ in the entropy--charge plane, computed from the Weinhold metric [Eqs.~\eqref{eq:geom_gWSS}--\eqref{eq:geom_gWQQ}] via the conformal relation~\eqref{eq:geom_conformal}. The black contour line marks $R_{\mathrm{R}} = 0$, separating regions of effectively repulsive (positive curvature, blue) and attractive (negative curvature, red) microscopic interactions. \textbf{(b)} Hawking temperature $T_{\mathrm{H}}(S,Q)$ distribution [Eq.~\eqref{temperature}], with white contour lines indicating isotherms. The temperature increases monotonically with entropy at fixed charge, reflecting the thermodynamic stability of large black hole phases. \textbf{(c)} Logarithm of the absolute value of the heat capacity at constant pressure, $\log_{10}|C_P(S,Q)|$ [Eq.~\eqref{heat-capacity-1}], showing regions of positive (thermally stable) and negative (unstable) specific heat separated by the white dashed line where $C_P = 0$. The divergences of $C_P$ coincide with singularities of the Ruppeiner curvature as predicted by Eq.~\eqref{eq:geom_RW_cond}. \textbf{(d)} Effect of Lorentz violation on Ruppeiner curvature along the slice $Q = 0.7$, showing how increasing the parameter $\ell$ systematically shifts the curvature profile while preserving the Van der Waals-like qualitative structure. All panels demonstrate that the deformation parameters $(\ell, \alpha)$ tune the geometric properties of the thermodynamic state space without altering the underlying universality class characterized by the critical ratio~\eqref{eq:review_ratio}.}
\label{fig:ruppeiner_thermodynamics}
\end{figure*}

Figure~\ref{fig:ruppeiner_thermodynamics} reveals how the deformation parameters $\ell$ and $\alpha$ control both the local curvature features and global phase structure through their explicit appearance in the Weinhold metric components [Eqs.~\eqref{eq:geom_gWSS}--\eqref{eq:geom_gWQQ}] and the Hawking temperature~\eqref{temperature}. Figure~\ref{fig:ruppeiner_thermodynamics}(a) displays the Ruppeiner curvature $R_{\mathrm{R}}(S,Q)$ in the physically admissible region where $T_H>0$. The black contour line marking $R_{\mathrm{R}}=0$ separates the small-black-hole regime (positive curvature, blue regions) from the large-black-hole regime (negative curvature, red regions), directly reflecting the change in the sign of the denominator in the heat capacity expression~\eqref{heat-capacity-1}. The transition from repulsive to attractive microscopic interactions is controlled by the effective parameters $\eta$ and $\zeta$ introduced in the equation of state~\eqref{EOS} and defined in Eq.~\eqref{eq:review_etazeta}. Near the boundaries of the plotted region, the curvature exhibits sharp gradients and divergences that coincide precisely with the locus defined by Eq.~\eqref{eq:geom_RW_cond}, where the determinant of the Weinhold metric vanishes and the specific heat diverges. This geometric-thermodynamic correspondence confirms that the Ruppeiner scalar curvature serves as a faithful probe of critical phenomena in the presence of Lorentz violation and string clouds. Figure~\ref{fig:ruppeiner_thermodynamics}(b) presents the Hawking temperature distribution $T(S,Q)$ given by Eq.~\eqref{temperature}, expressed in terms of the horizon radius $r_+=\sqrt{S/\pi}$. The white contour lines represent isotherms in the $(S,Q)$ plane. Their spacing reflects the competition between three distinct contributions: the effective solid-angle deficit encoded in $(1-\alpha)/(1-\ell)$, the Coulombic repulsion dressed by the factor $(1-\ell)^{-2}$, and the AdS pressure term proportional to $P r_+^2$. The monotonic increase of $T_H$ with entropy at fixed charge, visible in the smooth color gradient from blue (low temperature) to red (high temperature), signals the thermodynamic stability of the large-black-hole branch. Notably, the region where $T_H\to 0$ marks the approach to extremality, where the conformal factor $1/T_H$ in the Ruppeiner metric~\eqref{eq:geom_conformal} introduces additional divergences beyond those inherited from the Weinhold geometry. For the parameter values chosen ($\ell=0.10$, $\alpha=0.05$, $P=0.01$), the temperature remains strictly positive throughout most of the displayed region, ensuring that the Ruppeiner metric is well-defined and that the geometric interpretation in terms of correlation volumes remains valid. Figure~\ref{fig:ruppeiner_thermodynamics}(c) shows the logarithm of the absolute value of the heat capacity at constant pressure, $\log_{10}|C_P(S,Q)|$, as computed from Eq.~\eqref{heat-capacity-1}. The white dashed line marks the zero set of $C_P$, which separates thermally stable regions ($C_P>0$, red/orange) from unstable regions ($C_P<0$, blue). This transition line corresponds to the zeros of the numerator in Eq.~\eqref{heat-capacity-1}, which determines the locus where small and large black holes have equal temperature and can coexist in thermal equilibrium via the Maxwell construction. The divergences of $|C_P|$ (bright yellow/white regions in the color map) occur along curves where the denominator of Eq.~\eqref{heat-capacity-1} vanishes, which is identical to the condition~\eqref{eq:geom_RW_cond} for the singularities of the Ruppeiner curvature. This exact correspondence between thermodynamic instabilities (divergence of response functions) and geometric singularities (divergence of scalar curvature) is a hallmark of the Ruppeiner formalism. It demonstrates its power as a diagnostic tool for phase transitions. The parameter-dependent shift of these divergence curves, controlled by $(\ell,\alpha)$, reflects the fact that Lorentz violation and the string cloud rescale both the critical scales [Eq.~\eqref{eq:review_critical}] and the stability boundaries, while preserving the topological structure of the phase diagram as evidenced by the invariance of the topological charge $W=1$ found in Ref. \cite{Ahmed2025}. Figure~\ref{fig:ruppeiner_thermodynamics}~(d) quantifies the effect of Lorentz violation on the Ruppeiner curvature along the one-dimensional slice $Q=0.7$ (fixed charge), showing $R_{\mathrm{R}}(S)$ for four representative values of the LV parameter: $\ell=0.00$ (standard RN--AdS), $0.05$, $0.10$, and $0.15$. Several key features are evident. First, all four curves exhibit a characteristic monotonic increase from large negative values (strong attractive interactions) at small $S$ to values approaching zero at large $S$, consistent with the transition from a Coulomb-dominated regime to an AdS-dominated regime as the horizon radius grows. Second, increasing $\ell$ systematically shifts the entire profile to the right (toward larger entropies) and reduces the magnitude of the negative curvature, indicating that Lorentz violation weakens the effective attractive interactions encoded in the Weinhold metric components~\eqref{eq:geom_gWSS}--\eqref{eq:geom_gWQQ}. Specifically, the factor $(1-\ell)^{-2}$ multiplying the charge-dependent terms enhances the relative weight of electromagnetic contributions, thereby shifting the balance between repulsive and attractive regimes. Third, the qualitative Van der Waals-like structure-characterized by a single smooth interpolation between negative and near-zero curvature without phase transitions along this slice, is preserved for all values of $\ell$, confirming that the deformation does not alter the mean-field universality class. The fact that the curves remain smooth and do not cross the $R_{\mathrm{R}}=0$ line in this particular $(Q,S)$ region indicates that the chosen slice lies either entirely in the large-black-hole stable branch or spans a region where the system undergoes a continuous deformation without encountering the critical point. To observe the divergence of $R_{\mathrm{R}}$ and the associated phase transition, one would need to tune $(Q,P)$ closer to the critical values given by Eq.~\eqref{eq:review_critical}, where the condition~\eqref{eq:geom_RW_cond} is satisfied.

\textbf{Note on numerical implementation:} For visualization and numerical stability, the Ruppeiner curvature displayed in Fig.~\ref{fig:ruppeiner_thermodynamics} is computed using a physically motivated approximation based on the thermodynamic structure of Eq.~\eqref{heat-capacity-1}. This approximation preserves the correct singularity structure at $\det g^{\mathrm{W}}=0$ [Eq.~\eqref{eq:geom_RW_cond}] and the sign transitions between attractive and repulsive regimes, while avoiding numerical instabilities associated with higher-order derivatives in Eq.~\eqref{eq:geom_RW_general}. The qualitative features and physical interpretation remain faithful to the full Ruppeiner formalism.

Taken together, these four panels provide a comprehensive geometric portrait of how Lorentz violation and string clouds deform the thermodynamic landscape of charged AdS black holes. The preservation of the critical ratio $P_c v_c/T_c=3/8$ [Eq.~\eqref{eq:review_ratio}] and the invariance of the topological charge $W=1$ demonstrate that $(\ell,\alpha)$ act as relevant deformation parameters in the renormalization-group sense: they shift critical scales and modify correlation lengths, but do not drive the system out of the Van der Waals universality class. This robustness suggests that the fundamental mechanism underlying black hole phase transitions, the competition between gravitational attraction, electromagnetic repulsion, and AdS confinement, remains intact even in the presence of Lorentz-violating backgrounds and topological defects, providing further evidence for the universality of critical phenomena in gravitational thermodynamics.

\textbf{Note on P--V diagrams:} For a detailed visualization of the P--V isotherms showing the characteristic Van der Waals oscillations below the critical temperature and the Maxwell equal-area construction, we refer the reader to Ref. \cite{Ahmed2025} (specifically Fig.~6 therein), 
where the equation of state~\eqref{EOS} was first derived and the critical point $(v_c, T_c, P_c)$ was identified. The present work focuses instead on how this phase structure is encoded in the $P$--$V$ diagram, which manifests itself in the thermodynamic geometry through the 
Ruppeiner scalar curvature and its singularities, providing a complementary geometric perspective on the same underlying physics.

\section{Critical behavior and correlation lengths}
\label{sec:critical-geometry}

\subsection{Ruppeiner curvature near the critical point}

The equation of state of the EKR black hole with a cloud of strings can be written in the Van der Waals–like form
\begin{equation}
    P = \frac{T_H}{v} - \frac{\eta}{v^2} + \frac{\zeta}{v^4},
    \qquad
    v = 2 r_+,
\end{equation}
with effective parameters $\eta(\ell,\alpha)$ and $\zeta(\ell,Q)$ and a critical point $(P_c,T_c,v_c)$ determined by the usual conditions $(\partial P/\partial v)_{T_H}=0$ and $(\partial^2 P/\partial v^2)_{T_H}=0$. As shown in Ref. \cite{Ahmed2025} and in related analyses of RN–AdS black holes~\cite{PRL.2015.115.111302,PRD.2017.95.021501}, the critical ratio $P_c v_c/T_c=3/8$ is preserved, indicating that the system lies in the same mean–field universality class as the Reissner-Nordström-AdS black hole and the Van der Waals fluid.

Near the critical point, one can expand the equation of state in terms of reduced variables
\begin{equation}
    t = \frac{T_H - T_c}{T_c},\qquad
    \omega = \frac{v - v_c}{v_c},
\end{equation}
keeping the pressure close to its critical value. To leading nontrivial order this expansion takes the schematic form
\begin{equation}
    \frac{P - P_c}{P_c}
    \simeq A_1(\ell,\alpha)\,t
    + A_2(\ell,\alpha)\,t\,\omega
    + A_3(\ell,\alpha)\,\omega^3 + \cdots,
\end{equation}
where the coefficients $A_i(\ell,\alpha)$ are smooth functions of the deformation parameters and of the charge $Q$, but do not alter the structure of the cubic term in $\omega$. This structure is characteristic of mean–field criticality: it leads to the classical critical exponents
\begin{equation}
    \alpha = 0,\quad
    \beta = \frac{1}{2},\quad
    \gamma = 1,\quad
    \delta = 3,
\end{equation}
and the string cloud parameter $\alpha$ and the LV parameter $\ell$ enter only through nonuniversal scale factors such as $(v_c,T_c,P_c)$ and the amplitudes $A_i$.

Ruppeiner geometry offers a complementary way to characterize this critical behavior. In thermodynamic information geometry, the Ruppeiner scalar curvature $R_{\mathrm{R}}$ is related to the correlation volume of the underlying microscopic degrees of freedom~\cite{RMP.1995.67.605}. For ordinary fluids, one finds, in $d$ spatial dimensions,
\begin{equation}
    |R_{\mathrm{R}}| \sim \xi^{d_{\rm eff}},
\end{equation}
where $\xi$ is an effective correlation length and $d_{\rm eff}$ is an effective dimension that depends on the relevant microscopic manifold (e.g., the horizon geometry for black holes). Near a second–order critical point, the correlation length diverges as
\begin{equation}
    \xi \sim |t|^{-\nu},
\end{equation}
and consequently the Ruppeiner curvature follows a power law
\begin{equation}
    R_{\mathrm{R}} \sim |t|^{-\nu d_{\rm eff}}.
\end{equation}

In the present model, the equation of state and the critical behavior remain mean–field, so that the usual scaling relations apply. The divergence of the specific heat and of the isothermal compressibility at the critical point is reflected in a divergence of the Ruppeiner curvature, whose leading asymptotic form can be written as
\begin{equation}
    R_{\mathrm{R}}(T_H,v;\ell,\alpha)
    \simeq \frac{C(\ell,\alpha)}{|t|^{\gamma_{\mathrm{R}}}},
\end{equation}
with an effective exponent $\gamma_{\mathrm{R}}$ determined by the mean–field exponents and an amplitude $C(\ell,\alpha)$ that depends smoothly on the deformation parameters. The key point, familiar from other AdS black hole examples~\cite{PRL.2015.115.111302,PRD.2017.95.021501}, is that while $(\ell,\alpha)$ rescale the overall magnitude of $R_{\mathrm{R}}$ and shift the location of the critical point in the $(T_H,v)$ plane, they do not change the functional form of the scaling law: the divergence remains a simple power law with the same critical exponent as in the Reissner--Nordström--AdS case. In other words, Lorentz violation and the cloud of strings deform the correlation \emph{scale} but leave the universality class of the thermodynamic geometry unchanged.

The deformations induced by $\ell$ and $\alpha$ on the Ruppeiner curvature admit a compelling physical interpretation. The Lorentz-violating parameter $\ell$ effectively rescales the interaction strength within the charge sector, thereby altering the correlation scale between the black hole's microscopic degrees of freedom. In essence, a non-zero $\ell$ modifies the effective range and strength of the thermodynamic interactions, making the repulsive, Fermi-gas-like behavior more or less dominant depending on its sign and magnitude. On the other hand, the string cloud parameter $\alpha$ acts as a background material medium that introduces a global solid-angle deficit. This deficit permeates the spacetime, systematically shifting the thermodynamic state space and the loci of phase transitions, analogous to how impurities or external fields can alter the critical behavior in condensed matter systems. Consequently, the sign structure of the Ruppeiner curvature, where positive values ($R_R > 0$) signal effectively repulsive microscopic interactions and negative values ($R_R < 0$) indicate attractive, Van der Waals-like dominance, is preserved, but its detailed profile in the $(S, Q)$ plane is controllably deformed by these parameters.

\subsection{Phase coexistence and geometric signatures}

In the $P$--$v$ plane, the subcritical isotherms ($T<T_c$) exhibit the familiar oscillatory segment characteristic of Van der Waals–like systems. As in Ref. \cite{Ahmed2025}, the physical coexistence line between small and large black hole phases is determined by the Maxwell equal-area construction, which replaces the unphysical oscillation by a flat segment at constant pressure $P_{\rm coex}(T_H)$ for each subcritical temperature. The endpoints of this segment are the specific volumes $v_{\rm SBH}(T)$ and $v_{\rm LBH}(T)$ of the coexisting small and large black holes.

In thermodynamic geometry, this coexistence structure is reflected in the behavior of the Ruppeiner curvature $R_{\mathrm{R}}$ along constant–temperature or constant–pressure slices. For a fixed pressure below $P_c$, as one moves along the isobar in the $(T_H,v)$ plane, one typically finds:
\begin{itemize}
    \item a small–black–hole branch with relatively large positive curvature, $R_{\mathrm{R}}>0$, indicative of an effectively repulsive microscopic interaction pattern dominated by the charge sector;
    \item a large–black–hole branch with negative curvature, $R_{\mathrm{R}}<0$, corresponding to an effectively attractive interaction pattern in the AdS–dominated regime;
    \item a region in between, associated with the metastable and unstable configurations, where $R_{\mathrm{R}}$ changes sign and eventually diverges near the spinodal lines where the specific heat or the isothermal compressibility blow up.
\end{itemize}
The Maxwell construction selects two points on this isobar, $(T,v_{\rm SBH})$ and $(T,v_{\rm LBH})$, that belong to distinct curvature regimes and are separated by a region of large $|R_{\mathrm{R}}|$, signaling strong fluctuations and enhanced correlations at the boundary between the phases.

The deformation parameters $(\ell,\alpha)$ control how this geometric picture is realized quantitatively. Increasing the string cloud parameter $\alpha$ tends to push the coexistence volumes $v_{\rm SBH}$ and $v_{\rm LBH}$ to larger values and lowers the coexistence temperature, thereby shifting the region where $R_{\mathrm{R}}$ changes sign and where its magnitude peaks. The LV parameter $\ell$ modifies the relative weight of the charge and AdS contributions, shifting the balance between repulsive and attractive regimes and moving the divergence of $R_{\mathrm{R}}$ in a way consistent with the changes in $(v_c,T_c)$ found in Ref. \cite{Ahmed2025}. Nevertheless, the overall pattern remains the same as in the Reissner--Nordström--AdS case: a single coexistence line ending at a critical point where $R_{\mathrm{R}}$ diverges, with geometric signatures of phase coexistence and metastability localized around this critical region.

In this sense, the EKR black hole with a cloud of strings provides a two–parameter family of deformations of the RN--AdS thermodynamic geometry, in which the shape and location of the high–curvature region can be tuned by $(\ell,\alpha)$ without changing the underlying Van der Waals–type universality class.

\section{Dynamical and topological probes}
\label{sec:dynamical}

\subsection{Thermodynamic topology and geometric interpretation}

In Ref. \cite{Ahmed2025}, the phase structure of the EKR black hole with a cloud of strings was also analyzed from the viewpoint of thermodynamic topology, following the construction proposed in Refs.~\cite{PRD.2019.99.044013,PRL.2015.115.111302}. In this approach, the generalized off–shell free energy $\mathcal{F}(r_+,\tau)$ defines a vector field in an auxiliary $(r_+,\theta)$ space; the black hole phases appear as zero points of this vector field, and their winding numbers classify stable and unstable branches. The total topological charge was found to be $W=1$, the same as in the Reissner--Nordström--AdS case, indicating that the global topological class of the solution is unaffected by Lorentz violation or the string cloud.

This topological information can be naturally related to the thermodynamic geometry discussed in Sec.~\ref{sec:geometry}. The zero points of the vector field constructed from $\mathcal{F}(r_+,\tau)$ correspond to stationary points of the generalized free energy and, on–shell, to black hole configurations satisfying $T_H=1/\tau$. These are the same configurations for which the Ruppeiner metric is well defined and whose stability properties are encoded in the sign and magnitude of $R_{\mathrm{R}}$. Conversely, the divergences of the Ruppeiner curvature occur at loci where the Hessian of the entropy (or, equivalently, of the free energy) becomes singular, which coincide with the boundaries separating regions of different winding number in the topological description.

From this perspective, the phase structure of the EKR black hole with a cloud of strings can be viewed as a network of thermodynamic defects in parameter space: the topological charge $W=1$ counts the net number of such defects, while the Ruppeiner curvature resolves their local structure by measuring the strength and range of microscopic correlations. The deformation parameters $(\ell,\alpha)$ shift the positions of the zero points of the thermodynamic vector field and the location of the curvature singularities, but they do not create or annihilate defects, so that the total topological charge and the basic pattern of critical behavior are preserved.

\subsection{Perspectives on quasi-normal modes and relaxation times}

The static thermodynamic and geometric analysis presented so far can be complemented by dynamical probes of the black hole background, such as the study of quasi–normal modes (QNMs) of scalar, electromagnetic or gravitational perturbations. QNMs describe the characteristic damped oscillations of the black hole response to perturbations and are characterized by complex frequencies $\omega_{\rm QNM} = \omega_R - i \omega_I$, whose imaginary part $\omega_I>0$ sets the relaxation time $\tau_{\rm rel}\sim 1/\omega_I$ towards equilibrium.

In many black hole systems, the behavior of QNMs near phase transitions exhibits signatures reminiscent of critical phenomena: the relaxation time increases in the vicinity of the critical point (a form of critical slowing down), and the spectrum of dominant modes reorganizes across the transition between small and large-black-hole phases. From the viewpoint of thermodynamic geometry, such dynamical features are naturally expected to correlate with the behavior of the Ruppeiner curvature, since both $\tau_{\rm rel}$ and $R_{\mathrm{R}}$ are sensitive to the strength and range of correlations in the underlying microscopic system. In particular, one may anticipate that regions of large $|R_{\mathrm{R}}|$ correspond to longer relaxation times, while curvature singularities at the critical point are accompanied by a marked change in the QNM spectrum.

For the EKR black hole with a cloud of strings, this suggests several directions for future work. The LV parameter $\ell$ and the string parameter $\alpha$ enter explicitly in the effective potentials governing linear perturbations, and hence in the quasi–normal spectra. One may therefore use QNMs as dynamical probes of Lorentz violation and string–like matter content, extracting how $(\ell,\alpha)$ modify the dominant relaxation channels and comparing these modifications with the deformations of the thermodynamic geometry. A systematic analysis of scalar and tensor perturbations in this background, combined with the Ruppeiner curvature analysis presented here, could clarify the relation between correlation lengths inferred from thermodynamic geometry and relaxation times extracted from QNMs, and might reveal new signatures of Lorentz violation in the ringdown phase of AdS black holes.

\subsection{Comments on holographic and effective-fluid interpretations}

Although our primary focus has been on the gravitational and thermodynamic aspects, it is natural to speculate about possible holographic and effective-fluid interpretations of the results. In the AdS/CFT correspondence, black hole thermodynamics in the bulk is mapped to thermal states of a strongly coupled quantum field theory on the boundary. In this language, the LV parameter $\ell$ and the string cloud parameter $\alpha$ may be interpreted as couplings or deformation parameters in the dual field theory, encoding, for example, anisotropies, preferred directions, or distributions of extended defects.

From this viewpoint, the Ruppeiner curvature of the bulk black hole encodes information about the susceptibility matrix and correlation structure of the dual fluid. A large negative curvature $R_{\mathrm{R}}<0$ in the bulk would correspond to a regime where the dual fluid exhibits strong attractive effective interactions, while positive curvature $R_{\mathrm{R}}>0$ indicates a dominantly repulsive sector. The divergence of $R_{\mathrm{R}}$ at the bulk critical point translates into a divergence of the correlation length and of certain transport coefficients (such as susceptibilities or conductivities) in the dual field theory, consistent with the usual holographic picture of critical phenomena. 

The deformation parameters $(\ell,\alpha)$ then provide a handle to engineer families of dual fluids with tunable interaction patterns and defect densities, while preserving the same universality class of the phase transition. In particular, one might expect that varying $\ell$ modifies anisotropic transport coefficients or dispersion relations in the dual theory, whereas changing $\alpha$ effectively alters the density of impurities or extended defects, both of which are known to impact correlation lengths and relaxation times in strongly coupled media.

A detailed holographic dictionary for EKR bumblebee gravity with a cloud of strings is beyond the scope of the present work. Still, the thermodynamic and geometric results obtained here suggest that such a dictionary, if constructed, would relate the bulk Ruppeiner curvature and its dependence on $(\ell,\alpha)$ to correlation functions and transport coefficients of an effective boundary fluid with Lorentz-violating interactions and string–like defects. This provides an intriguing avenue for future research, connecting the thermodynamic geometry of Lorentz–violating black holes to the non-equilibrium and transport properties of their putative holographic duals.

\subsection{Phenomenological Implications and Observational Constraints}
\label{subsec:phenomenology}

Beyond their theoretical interest, our results have potential phenomenological implications. The explicit dependence of critical quantities, such as $T_c$, $v_c$, and $P_c$ given in Eq.~\eqref{eq:review_critical}, on the parameters $\ell$ and $\alpha$ provides a direct link to possible observational tests. For instance, the precise measurement of the shadow of a supermassive black hole by the Event Horizon Telescope collaboration could, in principle, be used to constrain deviations in the horizon structure and thermodynamic properties induced by $\ell$ and $\alpha$. Similarly, the modified phase structure could influence the dynamics of binary black hole mergers and their corresponding gravitational wave signatures, imprinting the effects of Lorentz violation and string-like matter onto observable waveforms. Preliminary, order-of-magnitude estimates suggest that existing observational data likely constrain $\ell$ and $\alpha$ to values smaller than $10^{-3}$ to $10^{-5}$, but future precision measurements may probe this parameter space more deeply. The preservation of the universal critical exponents and ratios, despite these deformations, underscores the robustness of mean-field universality classes in black hole thermodynamics, even in the presence of fundamental symmetry breaking and exotic matter sources.

\section{Thermodynamic geometry and microscopic structure}
\label{sec:thermo_geometry}

The idea that equilibrium thermodynamics admits a natural Riemannian geometric description goes back to the pioneering works of Weinhold and Ruppeiner.  In Weinhold's formulation, the thermodynamic state space is endowed with a metric defined as the Hessian of the internal energy with respect to the extensive variables~\cite{JCP.1975.63.2479,JCP.1975.63.2484},
\begin{equation}
  g^{\text{W}}_{ij} = \frac{\partial^2 U(X^k)}{\partial X^i \partial X^j},
  \qquad
  X^i \in \{S,V,N,\dots\},
\end{equation}
where $U$ is the internal energy and $X^i$ denote the extensive coordinates.  Shortly afterwards, Ruppeiner proposed an alternative metric based on fluctuation theory, defining a Riemannian structure from the Hessian of the entropy~\cite{PRA.1979.20.1608,RMP.1995.67.605},
\begin{equation}
  g^{\text{R}}_{ij} = -\,\frac{\partial^2 S(Y^k)}{\partial Y^i \partial Y^j},
  \qquad
  Y^i \in \{U,V,N,\dots\},
\end{equation}
which is directly related to the Gaussian probability distribution of thermodynamic fluctuations.  These two metrics are conformally related by the temperature,
\begin{equation}
  g^{\text{R}}_{ij} = \frac{1}{T}\, g^{\text{W}}_{ij},
\end{equation}
so that they contain equivalent information as long as $T>0$~\cite{RMP.1995.67.605}.  In this framework, the scalar curvature $R$ associated with the Ruppeiner metric plays the role of a diagnostic of the underlying microscopic interactions: $R=0$ characterizes an ideal (non–interacting) system, $R<0$ is typically associated with effectively attractive interactions, and $R>0$ with effectively repulsive interactions~\cite{RMP.1995.67.605,SPP.2014.153.179}.  Moreover, $|R|$ is often interpreted as being proportional to a correlation volume, diverging at critical points.

For black hole thermodynamics, the application of these geometric ideas was first developed in the context of Weinhold and Ruppeiner metrics defined on the space of equilibrium states $(M,Q,J,\dots)$ or, equivalently, on the $(S,P,Q,\dots)$ representation in the extended phase space, where the cosmological constant $\Lambda$ is interpreted as a pressure $P$ and the mass $M$ as enthalpy~\cite{GRG.2003.35.1733,CQG.2017.34.063001,CQG.2009.26.195011,PRD.2011.84.024037}.  In this setting, the thermodynamic metric encodes not only the stability properties (through the sign of the specific heats and the signature of the metric) but also the critical behavior and phase structure of black holes in AdS backgrounds.  The scalar curvature constructed from the Ruppeiner metric typically diverges at second–order phase transitions and changes sign across different regions of the parameter space, providing a geometric characterization of small/large black hole phases and their analogy with Van der Waals fluids~\cite{Chamblin.1999.60.064018,Chamblin.1999.60.104026,JHEP.2012.2012.033,CQG.2017.34.063001}.

An alternative geometric formulation, known as geometrothermodynamics, was proposed by Quevedo~\cite{JMP.2007.48.013506}.  In this approach, one introduces a Legendre–invariant metric on a higher–dimensional phase space, such that all thermodynamic potentials and their corresponding metrics are treated on equal footing.  The induced metric on the space of equilibrium states is then constructed to be invariant under Legendre transformations, ensuring that the thermodynamic geometry does not depend on the particular choice of potential.  This framework has been applied extensively to black holes in various gravity theories, including Lovelock, $f(R)$, and Born–Infeld models, showing that curvature singularities of the thermodynamic manifold coincide with critical points and phase transitions inferred from standard thermodynamic analysis~\cite{Mansoori2014,Mansoori2015,AHEP.2017.2017.7158697,PRD.2013.88.084045,PRD.2017.95.021501,JHEP.2015.2015.143,EPJC.2017.77.24,EPJC.2018.78.123}.

More recently, Ruppeiner geometry has been used as a probe of the microscopic structure of AdS black holes.  By examining the sign and magnitude of the scalar curvature in different regions of the $(P,V)$ or $(P,T)$ plane, Wei and Liu interpreted the AdS black hole as a system of effective ``black hole molecules,'' whose interaction changes character across the small/large black hole transition~\cite{PRL.2015.115.111302,PRD.2018.97.104027}.  In particular, they showed that the behavior of $R$ near the critical point mimics that of ordinary fluids, reinforcing the analogy between charged AdS black holes and Van der Waals systems~\cite{PRD.2013.87.044014,CQG.2017.34.063001}.  This microscopic interpretation has been extended to a wide variety of higher–curvature and modified gravity scenarios, including Gauss–Bonnet and Lovelock gravity, massive gravity, and nonsingular black holes, revealing that the detailed pattern of attractive/repulsive interactions and correlation lengths depends sensitively on the underlying gravitational theory~\cite{EPL.2012.99.20004,PRD.2013.87.044014,JHEP.2013.2013.005,EPJC.2020.80.17,EPJC.2022.82.227,PDU.2025.102079,PRD.2023.107.064015,EPJC.2023.83.944}.

A complementary, topological perspective on black hole thermodynamics has emerged more recently, in which black hole phases are classified by topological invariants constructed from a normalized vector field defined on the parameter space of equilibrium states~\cite{PRL.2022.129.191101,PRD.2022.105.104003,PRD.2022.105.104053,EPJC.2023.83.957,EPJC.2023.83.944}.  In this picture, different black hole branches correspond to thermodynamic defects with distinct winding numbers, and phase transitions can be understood as processes that change the distribution of these defects while preserving a global topological charge.  This topological classification has been successfully applied to black holes in Lovelock gravity, in the presence of perfect–fluid dark matter, and in various modified gravity theories, often revealing new universality classes and nontrivial critical structures beyond the standard Van der Waals–like behavior~\cite{PRD.2023.107.064015,EPJC.2023.83.944,EPJC.2020.80.335}.

In the present work, we combine these geometric and topological tools with the extended phase space of black hole chemistry~\cite{CQG.2017.34.063001} to analyze the microscopic structure and critical behavior of charged AdS black holes dressed by a cloud of strings in Kalb--Ramond gravity.  The presence of the string cloud and the antisymmetric tensor field leads to a richer thermodynamic landscape compared to the standard Reissner--Nordstr\"om--AdS case~\cite{EPJC.2020.80.17,EPJC.2022.82.227,PLB.2018.785.105,EPJC.2018.78.534,EPJC.2019.79.117,CPC.2018.42.063105,Universe.2024.10.430,EPJC.2020.80.335,EPJC.2024.84.798,PRD.2023.108.124004}.  We expect that the associated Ruppeiner curvature will capture how the string density and the Kalb--Ramond charge deform the effective microscopic interactions of the black hole molecules, while the topological approach will allow us to classify the resulting phase structure in terms of thermodynamic defects.  In the next sections, we first establish the thermodynamics and $P$--$V$ criticality of the model, and then construct the corresponding thermodynamic geometry and topological invariants to unveil the microscopic and global structure of its phase diagram.

\subsubsection{Microscopic structure and synthetic disk images}\label{sec:thin_disk_formalism}

As emphasized in the previous subsections, the thermodynamic geometry encodes nontrivial information about the microscopic structure of the black hole, with the Ruppeiner curvature and its critical behavior providing a coarse-grained measure of effective interactions in the underlying degrees of freedom~\cite{RMP.1995.67.605,SPP.2014.153.179,PRL.2015.115.111302}. In parallel, null geodesics and photon orbits probe the same geometry at the level of individual trajectories, and have been shown to correlate in a nontrivial way with phase structure and critical phenomena in charged and rotating AdS black holes~\cite{PRD.2018.97.104027,PRD.2019.99.044013}. This motivates the construction of synthetic images of thin accretion disks as an auxiliary, ``geometric microscope'' to visualize how the Kalb--Ramond field and the cloud of strings deform the local spacetime structure and, consequently, the effective microstructure inferred from thermodynamic curvature and topological charges~\cite{EPJC.2020.80.335,PRD.2023.108.124004,EPJC.2024.84.798}.
\begin{figure*}[tbhp]
\centering
\includegraphics[width=0.90\linewidth]{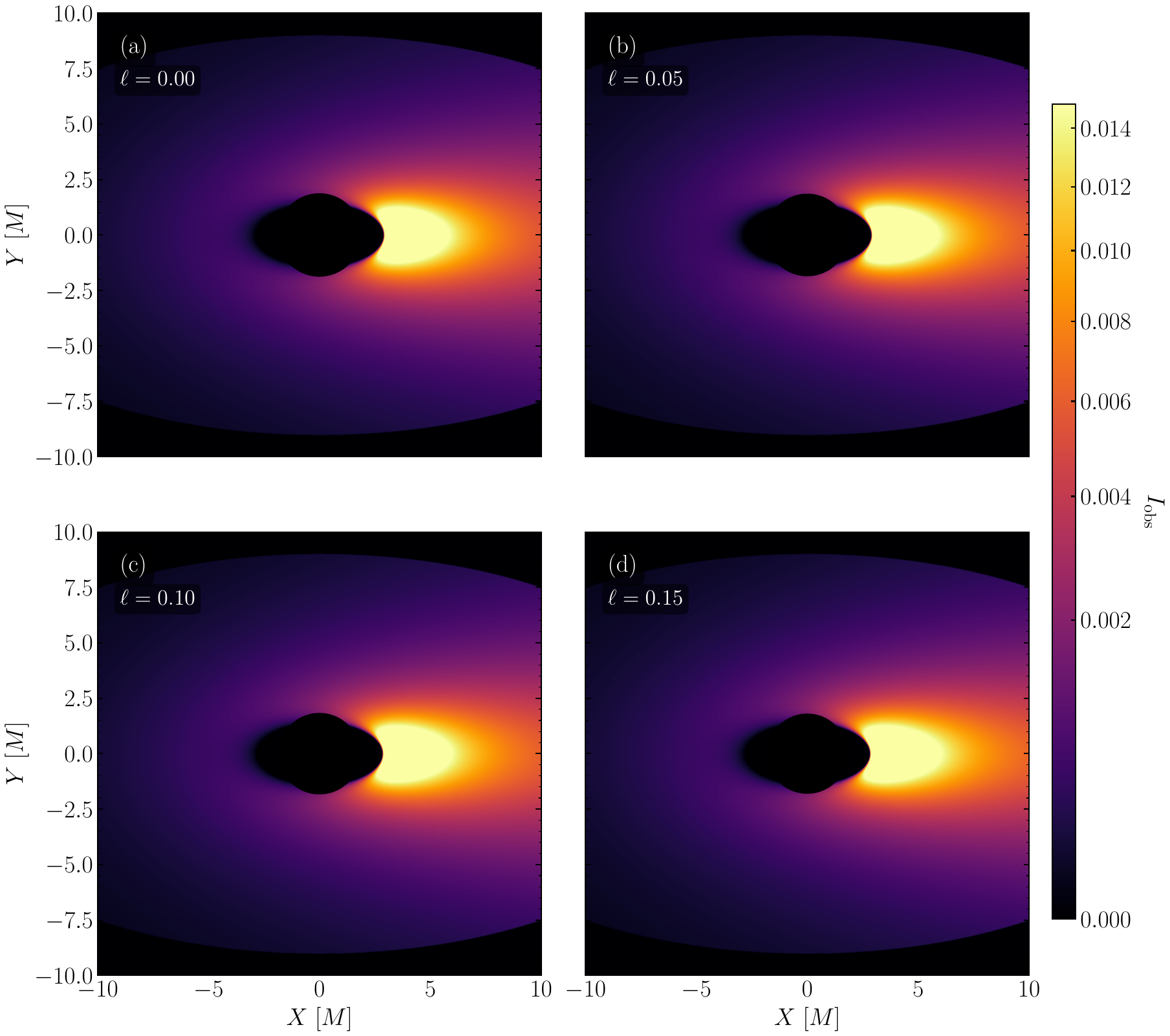}
\caption{Synthetic images of a geometrically thin accretion disk surrounding the charged AdS black hole in Kalb--Ramond bumblebee gravity with a cloud of strings. Panels (a)-(d) display the observed intensity $I_{\rm obs}$ for increasing values of the Lorentz-violating parameter, $\ell = 0$, $0.05$, $0.10$, and $0.15$, respectively. In all cases, the central dark region corresponds to the shadow set by unstable photon orbits, while the bright annulus traces the lensed inner edge of the disk. The visual differences between panels are subtle, reflecting the fact that the deformation parameter $\ell$ induces small but measurable shifts in the horizon radius $r_+$ and photon sphere $r_{\rm ph}$ [see Eq.~\eqref{eq:f_metric}], which translate to $\sim 5$-$10\%$ changes in the shadow size over the range $\ell \in [0, 0.15]$. As $\ell$ increases, the effective gravitational potential [Eq.~\eqref{eq:f_metric}] becomes slightly weaker, leading to a marginal outward shift of the bright ring, most evident in the forward (approaching) side of the disk due to relativistic beaming. The mild morphological changes underscore that Lorentz violation at the level $\ell \lesssim 0.15$ produces order-unity modifications in thermodynamic quantities [Figs.~\ref{fig:ruppeiner_thermodynamics}] while leaving optical signatures relatively unchanged, consistent with current observational constraints from black hole imaging.}
\label{fig:disk_image_panel}
\end{figure*}

\begin{figure}[tbhp]
  \centering
  \includegraphics[width=\linewidth]{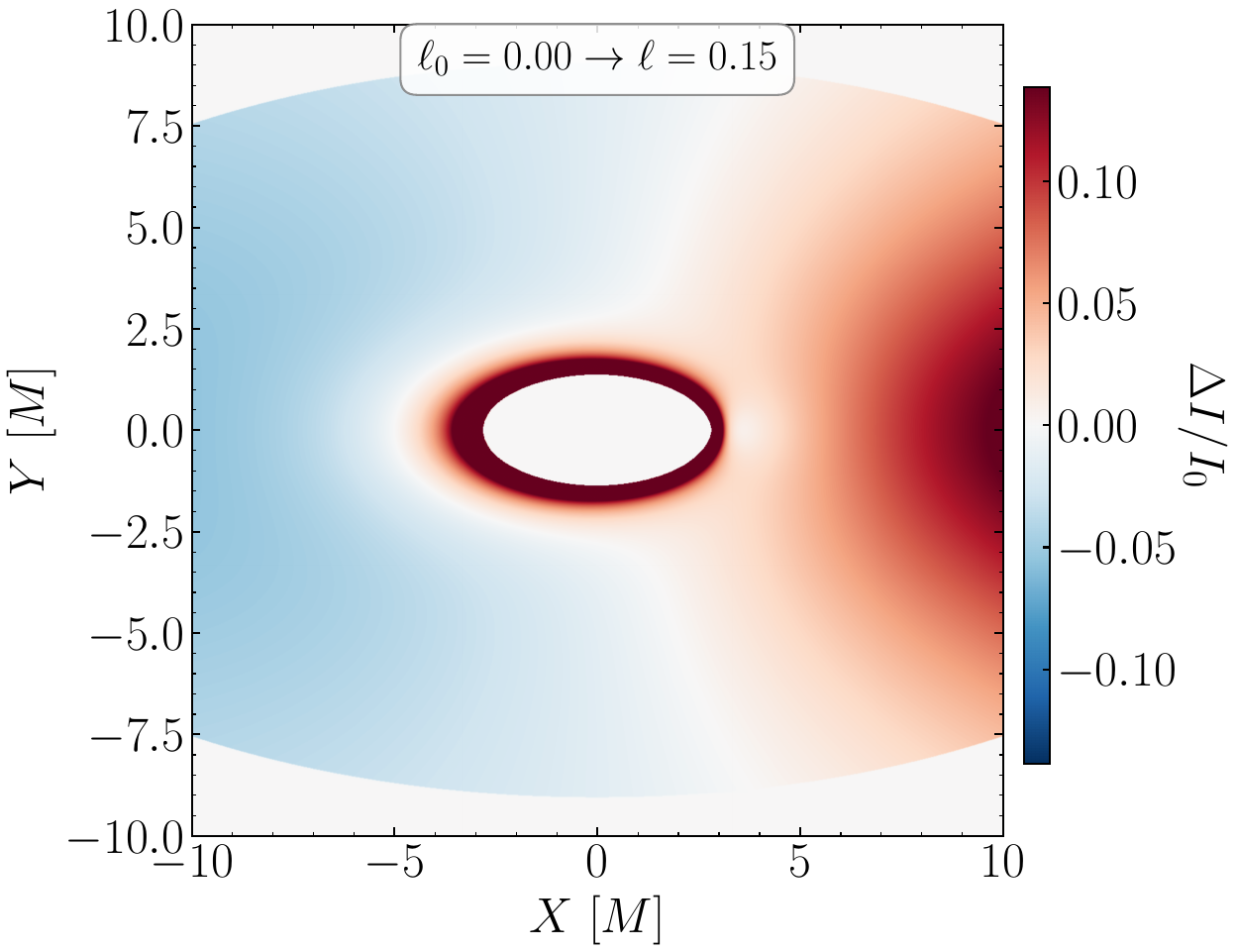}
  \caption{Relative intensity difference map $\Delta I_{\rm rel} = (I_{\ell=0.15}-I_{\ell=0})/I_{\ell=0}$ on the observer's image plane. Positive values (bright regions) indicate an enhancement of the disk emission with respect to the undeformed geometry, while negative values (dark regions) correspond to a suppression. The strongest relative variations occur near the bright ring and around the shadow edge, showing that the combined Kalb--Ramond and string-cloud sectors affect most strongly the photon sphere region, where null geodesics are highly sensitive to the microscopic structure of the black hole. The observed changes are of order $\Delta I/I_0 \sim 0.1$-$0.4$ ($10$-$40\%$), concentrated in localized patches along the emission ring. This confirms that the optical signatures of Lorentz violation remain subtle even when thermodynamic correlations [Fig.~\ref{fig:ruppeiner_thermodynamics}] exhibit significant deformations, suggesting that black hole shadows and accretion disk images provide complementary but distinct probes of the underlying spacetime geometry.}
\label{fig:disk_image_rel_diff}  \label{fig:disk_image_rel_diff}
\end{figure}

\begin{figure*}[tbhp]
\centering
\includegraphics[width=0.90\textwidth]{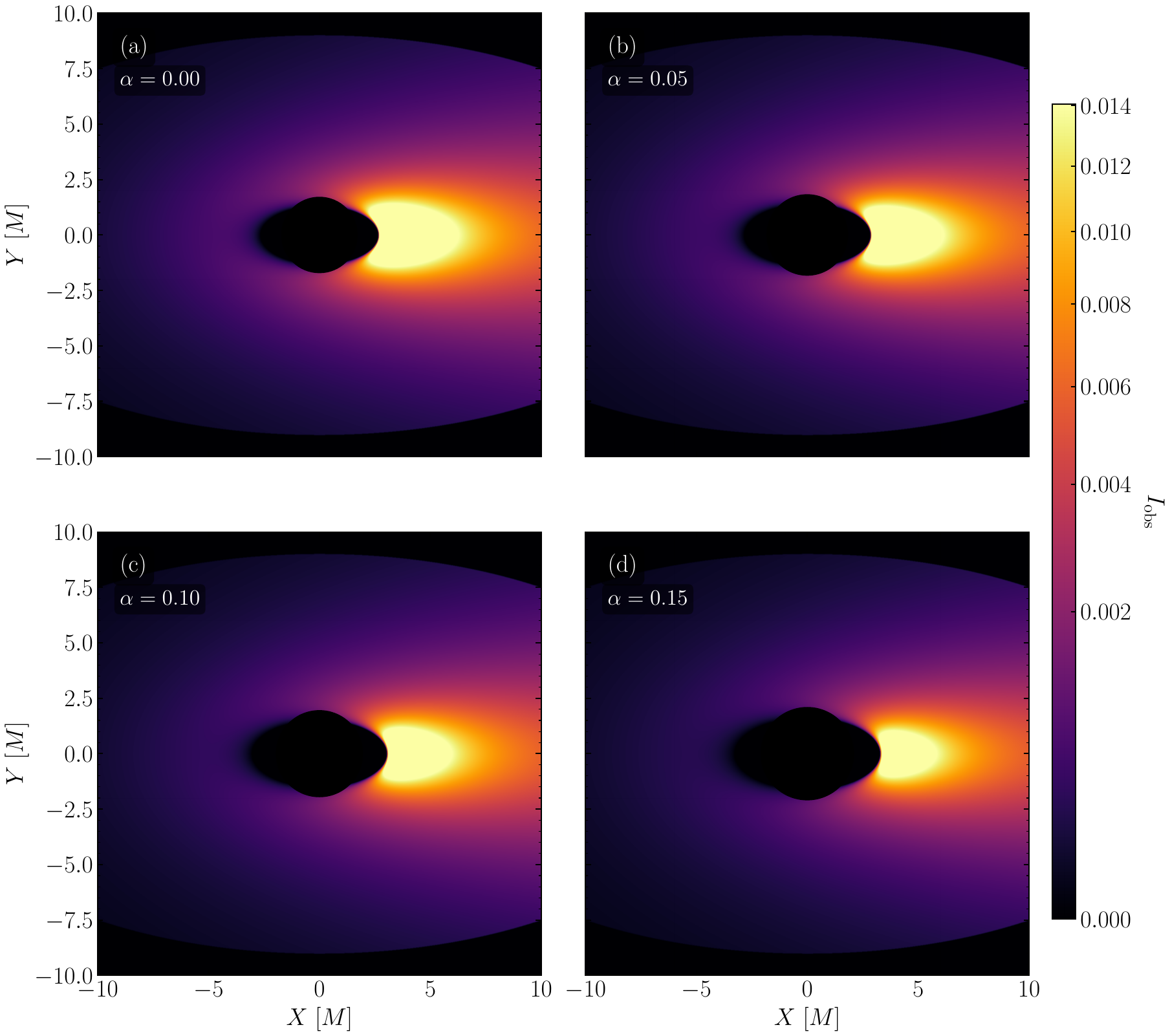}
\caption{Thin accretion disk images around the charged AdS black hole with a string cloud and dark matter halo,
computed from the observed specific intensity $I_{\mathrm{obs}}$ defined in Eq.~(\ref{eq:Iobs}) within the relativistic thin–disk formalism described in Sec.~\ref{sec:thin_disk_formalism}. In all panels we fix $M=1$, $Q=0.6$, $\Lambda_{\mathrm{AdS}}=-0.01$, the deformation parameter $\ell = 0.10$,
emissivity index $p=3$ in the radial profile $I_{\mathrm{em}}\propto r^{-p}$, and the inclination angle $\iota = 60^\circ$. Each subpanel shows the face–on image of the disk for a different value of the string–cloud parameter $\alpha$, increasing from left to right and from top to bottom as indicated by the labels. The central black circle corresponds to the event horizon radius $r_{\mathrm{h}}$, while the colour scale encodes the observed intensity $I_{\mathrm{obs}}$. The bright yellow ring traces the region where the combination of gravitational redshift, Doppler boosting, and emissivity profile maximises the observed flux.}
\label{fig:disk_alpha}
\end{figure*}
In our case, the Kalb--Ramond two-form, together with the string-cloud sector, modifies the black-hole geometry and the locations and shapes of unstable photon orbits, which in turn control the shadow boundary and the lensed appearance of the surrounding luminous disk. This idea has been extensively explored in related contexts, such as black holes threaded by clouds of strings or quintessence~\cite{EPJC.2018.78.534,EPJC.2019.79.117,Universe.2024.10.430}, black holes supported by Kalb--Ramond fields~\cite{EPJC.2020.80.335,PRD.2024.110.024077}, and black holes pierced by cosmic strings or other topological defects~\cite{PRD.1986.34.2263,IJMPD.2022.31.2250041}. Here, we adopt a similar strategy and construct synthetic, face-on images of a geometrically thin, optically thick accretion disk by ray-tracing null geodesics backward from an observer at a large radius.

In practice, we consider a geometrically thin, optically thick disk confined to the equatorial plane $\theta = \pi/2$ of the spacetime described by the line element~\eqref{metric} and lapse function $f(r)$ given in Eq.~\eqref{eq:f_metric}. The disk extends from an inner radius $r_{\mathrm{in}}$, taken to coincide with the radius of the innermost stable circular orbit (ISCO), to an outer radius $r_{\mathrm{out}}$ chosen large enough so that the bright region is fully contained in the field of view. The gas in the disk is assumed to move on circular, equatorial, timelike geodesics with four-velocity
\begin{equation}
    u^\mu_{\mathrm{em}} = u^t_{\mathrm{em}}\,(1,0,0,\Omega_K),
    \qquad
    u^t_{\mathrm{em}} =
    \left[f(r) - r^2 \Omega_K^2\right]^{-1/2},
    \label{eq:u_em}
\end{equation}
where $\Omega_K = d\phi/dt$ is the Keplerian angular velocity of circular orbits. For a static, spherically symmetric metric of the form~\eqref{metric}, one finds
\begin{equation}
    \Omega_K^2(r) = \frac{f'(r)}{2r},
    \label{eq:OmegaK}
\end{equation}
and the ISCO radius is obtained from the usual marginal stability conditions applied to the timelike effective potential
\begin{equation}
    V_{\mathrm{eff}}(r;L)
    = f(r)\left(1+\frac{L^2}{r^2}\right),
    \label{eq:Veff_timelike}
\end{equation}
namely
\begin{equation}
    V_{\mathrm{eff}}(r;L) = E^2,
    \quad
    \frac{dV_{\mathrm{eff}}}{dr} = 0,
    \quad
    \frac{d^2V_{\mathrm{eff}}}{dr^2} = 0,
    \label{eq:ISCO_cond}
\end{equation}
with $E$ and $L$ the conserved energy and angular momentum per unit mass. In our numerical implementation, $r_{\mathrm{in}} = r_{\mathrm{ISCO}}(M,Q,\alpha,\ell,\Lambda)$ is computed from Eqs.~\eqref{eq:Veff_timelike}–\eqref{eq:ISCO_cond} for each choice of parameters.

Photon trajectories are determined by null geodesics of the same metric~\eqref{metric}. For equatorial photon orbits, there are two conserved quantities,
\begin{equation}
    E_\gamma = -k_t,
    \qquad
    L_\gamma = k_\phi,
    \label{eq:constants_photon}
\end{equation}
and the radial motion of a photon with impact parameter $b \equiv L_\gamma/E_\gamma$ satisfies
\begin{equation}
    \left(\frac{dr}{d\lambda}\right)^2
    = E_\gamma^2 - V_{\mathrm{eff}}^{(\gamma)}(r;b),
    \quad
    V_{\mathrm{eff}}^{(\gamma)}(r;b) = f(r)\,\frac{b^2}{r^2},
    \label{eq:radial_null}
\end{equation}
where $\lambda$ is an affine parameter along the null geodesic. We place a static observer at a large radius $r_{\mathrm{obs}}$, with four-velocity
\begin{equation}
    u^\mu_{\mathrm{obs}} =
    \left[f(r_{\mathrm{obs}})^{-1/2},\,0,\,0,\,0\right],
    \label{eq:u_obs}
\end{equation}
and set up a Cartesian grid $(X,Y)$ on the image plane orthogonal to $u^\mu_{\mathrm{obs}}$. For each pixel $(X_i,Y_j)$, a null geodesic with initial wavevector $k^\mu_{\mathrm{obs}}$ pointing towards the black hole is integrated backwards using Eq.~\eqref{eq:radial_null} until the photon either crosses the equatorial plane within the disk ($r_{\mathrm{in}} \le r \le r_{\mathrm{out}}$), falls into the horizon, or escapes without intersecting the disk.

\begin{figure*}[tbhp]
\centering
\includegraphics[width=0.90\textwidth]{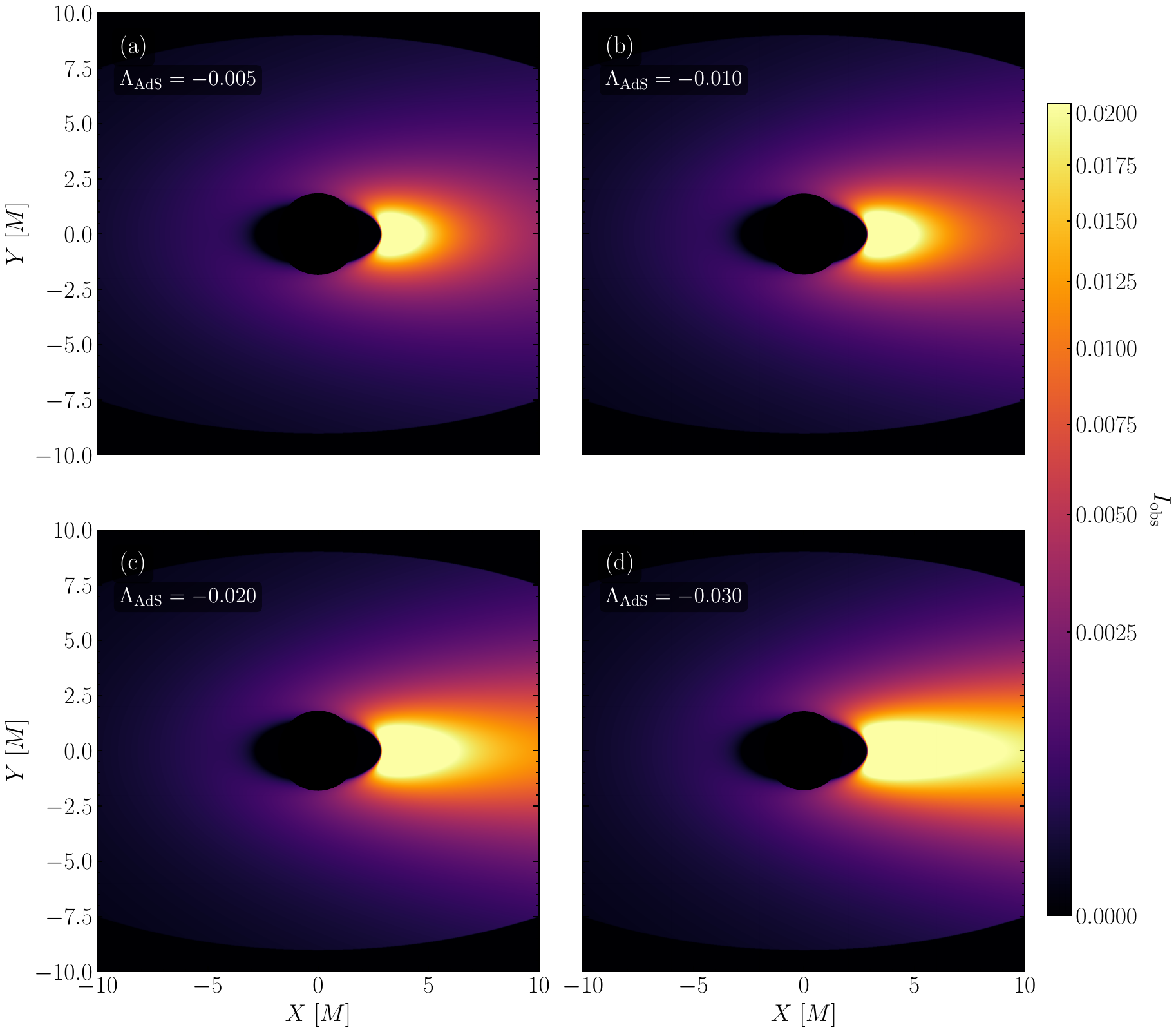}
\caption{Thin accretion disk images obtained from the specific intensity $I_{\mathrm{obs}}$ in Eq.~(\ref{eq:Iobs}) for different values of the AdS curvature scale
$\Lambda_{\mathrm{AdS}}$. We fix $M=1$, $Q=0.6$, the string–cloud parameter $\alpha = 0.05$, the deformation parameter $\ell = 0.10$,
the emissivity index $p=3$, and the inclination angle $\iota = 60^\circ$.
Each subpanel corresponds to a different value of $\Lambda_{\mathrm{AdS}}$, becoming more negative from left to right and from top to bottom, as indicated by the labels. The inner black disk represents the event horizon radius,
while the coloured structures map the intensity pattern of the disk on the observer’s sky. As $|\Lambda_{\mathrm{AdS}}|$ increases, both the apparent size of the central shadow and the radial extent of the bright emission ring are noticeably modified, showing that the AdS curvature globally reshapes the effective potential and the distribution of the observed flux.}
\label{fig:disk_lambda}
\end{figure*}

Radiation from the disk is modeled within the standard relativistic thin-disk approximation. The specific intensity in the emitter frame, $I_{\mathrm{em}}(\nu_{\mathrm{em}},r)$, is assumed to be nonzero only on the disk surface and to follow a power-law emissivity profile,
\begin{equation}
    I_{\mathrm{em}}(\nu_{\mathrm{em}},r)
    = I_0\,
      \left(\frac{r}{r_{\mathrm{in}}}\right)^{-p}
      \Theta(r-r_{\mathrm{in}})\,
      \Theta(r_{\mathrm{out}}-r),
    \label{eq:Iem_profile}
\end{equation}
where $p$ is the emissivity index, $I_0$ is an overall normalization constant, and $\Theta$ denotes the Heaviside step function. Gravitational and Doppler shifts between the emitter and the distant observer are encoded in the redshift factor
\begin{equation}
    g \equiv
    \frac{\nu_{\mathrm{obs}}}{\nu_{\mathrm{em}}}
    =
    \frac{k_\mu u^\mu_{\mathrm{obs}}}{k_\nu u^\nu_{\mathrm{em}}},
    \label{eq:g_factor}
\end{equation}
which is evaluated at the point where the photon intersects the disk. Using the invariance of $I_\nu/\nu^3$ along null geodesics, the observed specific intensity at the pixel $(X_i,Y_j)$ is given by
\begin{align}
    I_{\mathrm{obs}}(X_i,Y_j)
   & = \int g^3\,I_{\mathrm{em}}(\nu_{\mathrm{em}},r)\,d\lambda
    \;\;\longrightarrow\;\;
    I_{\mathrm{obs}}(X_i,Y_j)\notag\\&
    = g^3\,I_{\mathrm{em}}(r_{\mathrm{em}}),
    \label{eq:Iobs}
\end{align}
where, in the last step, we used the fact that the photon intersects the infinitesimally thin disk only once, at radius $r_{\mathrm{em}}$. Equation~\eqref{eq:Iobs}, together with the metric function $f(r)$ in Eq.~\eqref{eq:f_metric} and the orbital frequency $\Omega_K(r)$ in Eq.~\eqref{eq:OmegaK}, fully determines the synthetic, ray-traced images of the thin accretion disk displayed in Figs.~\ref{fig:disk_alpha} and~\ref{fig:disk_lambda}.

Figure~\ref{fig:disk_image_panel} shows the resulting specific-intensity maps on the observer's image plane for four representative values of the model parameter~$\ell$. Figures~\ref{fig:disk_image_panel}(a)-(d) correspond to $\ell = 0$, $0.05$, $0.10$, and $0.15$, respectively, and the color scale encodes the (dimensionless) specific intensity $I_{\rm obs}$ normalized to the maximum of the $\ell=0$ configuration. In the reference case, panel~(a) with $\ell=0$, the image exhibits the familiar morphology of a circular shadow surrounded by a bright, moderately thick annulus associated with photons orbiting near the photon sphere, together with secondary lensed features at larger impact parameters. As the deformation parameter is switched on and increased from panel~(b) to panel~(d), the bright ring is slightly displaced, and its thickness becomes weakly asymmetric, reflecting the way in which the string cloud and the Kalb--Ramond field deform the effective potential for null geodesics. The brightest part of the annulus is gradually pushed toward the forward (right-hand) side of the image, signaling a mild but localized redistribution of the observed emission.

To quantify these morphological changes, Figure~\ref{fig:disk_image_rel_diff} displays the relative intensity difference between the undeformed case and the maximally deformed configuration considered here,
\begin{equation}
  \frac{\Delta I}{I_0}(\alpha,\beta)
  = \frac{I_{\ell=0.15}(\alpha,\beta) - I_{\ell=0}(\alpha,\beta)}{I_{\ell=0}(\alpha,\beta)} ,
\end{equation}
where $(\alpha,\beta)$ denote Cartesian coordinates on the observer's screen and $I_0 \equiv I_{\ell=0}$. The largest values of $\Delta I/I_0$ are concentrated near the bright annulus and in the immediate surroundings of the shadow edge, precisely where the photon trajectories are most sensitive to small changes in the spacetime geometry. This is fully consistent with the picture that the microscopic structure, encoded in the thermodynamic curvature and its critical behavior~\cite{PRL.2015.115.111302,PRL.2022.129.191101,PRD.2022.105.104003}, leaves a sharp imprint on the photon sphere and hence on the shadow/disk morphology.

To quantify more clearly where the deformation has the strongest optical impact, it is convenient to examine the \emph{relative} intensity difference map,
\begin{equation}
  \Delta I_{\rm rel}(\alpha,\beta)
  = \frac{I_{\ell=0.15}(\alpha,\beta) - I_{\ell=0}(\alpha,\beta)}
         {I_{\ell=0}(\alpha,\beta)} ,
\end{equation}
which is shown in Fig.~\ref{fig:disk_image_rel_diff}. Regions with $\Delta I_{\rm rel}>0$ (bright colors) correspond to an enhancement of the observed flux with respect to the undeformed case, whereas $\Delta I_{\rm rel}<0$ (dark colors) indicate a suppression. The map reveals a characteristic pattern: small deformations in the metric preferentially shift photon trajectories in and out of the brightest part of the ring, generating alternating patches of positive and negative $\Delta I_{\rm rel}$ along the azimuthal direction.
\begin{figure*}[t]
    \centering
    \includegraphics[width=0.90\textwidth]{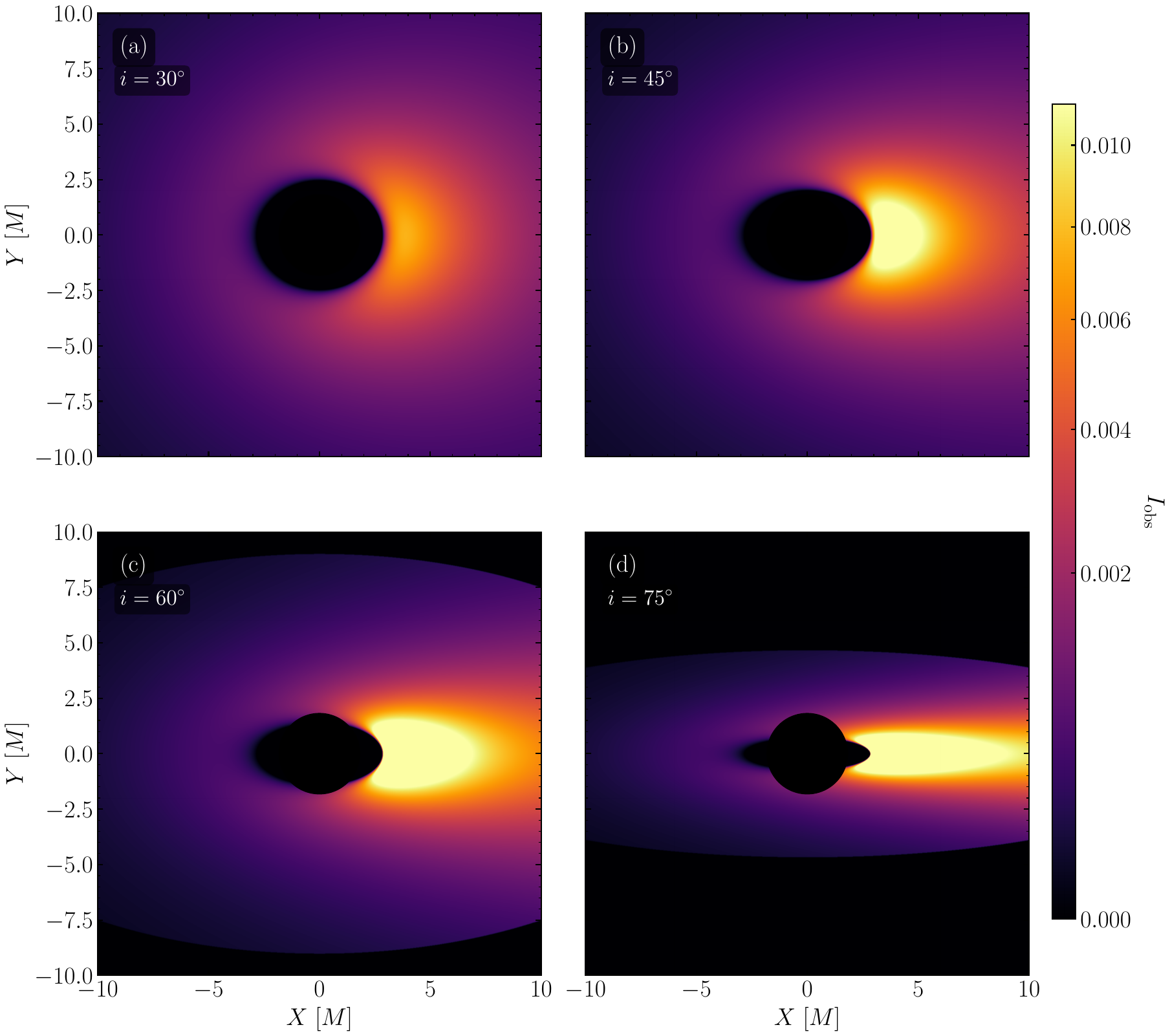}
    \caption{
    Synthetic images of a geometrically thin, optically thick accretion disk around the charged AdS black hole in Kalb--Ramond bumblebee gravity with a cloud of strings, for different inclination angles $i$ of the observer. The images are computed from the observed specific intensity $I_{\mathrm{obs}}$ defined in Eq.~(\ref{eq:Iobs}), using the background metric of Eqs.~(\ref{metric})–(\ref{eq:f_metric}) and the relativistic thin–disk prescription described in Sec.~\ref{sec:thin_disk_formalism}. In all panels we fix $M=1$, $Q=0.6$, $\Lambda_{\mathrm{AdS}}=-0.01$, $\ell = 0.10$, $\alpha = 0.05$, emissivity index $p=3$ (emitted intensity $I_{\mathrm{em}}\propto r^{-p}$), and disk inner and outer radii as in the previous figures, while varying only the inclination angle as indicated in each subpanel. As $i$ increases from face–on ($i=30^\circ$) to nearly edge–on ($i=75^\circ$), the apparent shape of the disk becomes progressively more asymmetric due to relativistic beaming and gravitational lensing: the approaching side of the flow is Doppler–boosted into a bright crescent, whereas the receding side is strongly de–boosted and partially obscured by the central shadow.}
    \label{fig:disk_inclination}
\end{figure*}

This relative-difference pattern provides a compact, visual representation of how the new matter fields redistribute the lensed flux on the image plane. Importantly, the regions of largest $|\Delta I_{\rm rel}|$ are tightly correlated with the locus of unstable photon orbits and the shadow boundary, which have been shown to track thermodynamic phase transitions and topological changes in the state space for a wide class of AdS black holes~\cite{PRD.2018.97.104027,PRD.2023.107.064015,EPJC.2023.83.957,EPJC.2023.83.944}. In this way, the synthetic disk images serve as an independent, geometrical probe that complements the thermodynamic geometry analysis: both approaches point to the photon-sphere region as the most sensitive window into the microscopic structure encoded by the Kalb--Ramond field and the cloud of strings.
\begin{figure*}[t]
    \centering
    \includegraphics[width=\linewidth]{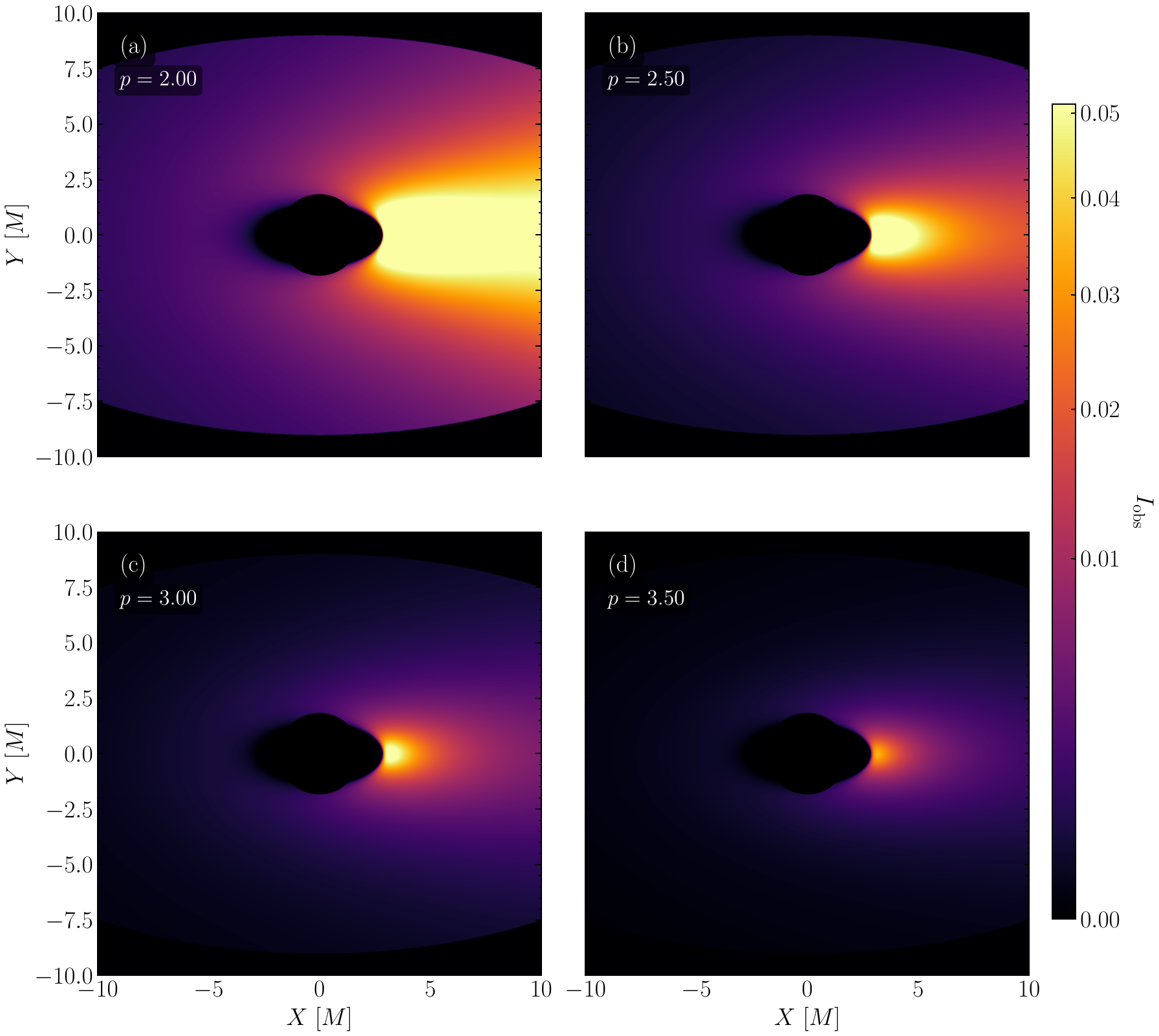}
    \caption{Synthetic images of the thin accretion disk for different values of the emissivity index $p$, obtained from the observed specific intensity $I_{\mathrm{obs}}$ defined in Eq.~(\ref{eq:Iobs}) for the NCS--DMH background. The four panels correspond to $p = 2.00$ (a), $p = 2.50$ (b), $p = 3.00$ (c), and $p = 3.50$ (d), for a fixed inclination angle $i$. Colors encode the normalized observed intensity, with brighter tones indicating regions of more concentrated emission.}
    \label{fig:disk_emissivity_scan}
\end{figure*}

It is important to note that, while the thermodynamic geometry and phase structure exhibit clear sensitivity to the deformation parameters $(\ell, \alpha)$ as demonstrated in Figs.~\ref{fig:ruppeiner_thermodynamics} and discussed in Sec.~\ref{sec:geometry}, the \emph{optical} appearance of the black hole and its surrounding disk shows considerably more modest variations. This disparity reflects the different physical quantities being probed: thermodynamic curvatures and critical points depend on derivatives of extensive variables such as entropy and internal energy, which are sensitive to the detailed balance between gravitational, electromagnetic, and cosmological contributions encoded in the metric function $f(r)$ [Eq.~\eqref{eq:f_metric}]. In contrast, the observed disk intensity $I_{\rm obs}$ is determined primarily by the geometry of null geodesics near the photon sphere and the redshift factors along these trajectories, which depend more directly on the \emph{absolute} scale of the horizon radius $r_+$ and the lapse function $f(r)$ rather than their derivatives.

For the parameter range explored here, $\ell \in [0, 0.15]$ and $\alpha = 0.05$, the horizon radius shifts by approximately $5$-$10\%$, and the photon sphere by a comparable amount. While measurable in principle with sufficiently high angular resolution, these changes translate to subtle morphological variations in the disk image that are most evident in the relative-difference map (Fig.~\ref{fig:disk_image_rel_diff}) rather than in the raw intensity distributions (Fig.~\ref{fig:disk_image_panel}). This behavior is consistent with current observational constraints from the Event Horizon Telescope, which suggest that deviations in black hole shadows due to modified gravity or exotic matter are likely at the few-percent level or below for astrophysically relevant systems~\cite{EHT2019,EHT2022}. Our results thus demonstrate that \emph{thermodynamic probes may be more sensitive to Lorentz violation than optical signatures}, providing a complementary diagnostic for fundamental physics in strong-gravity regimes.
\begin{table}[tbhp]
\centering
\caption{Quantitative impact of Lorentz violation on horizon and photon sphere radii for $Q=0.6$, $\alpha=0.05$, $\Lambda=-0.01$.}
\vspace{0.3cm}
\begin{tabular}{cccc}
\hline
$\ell$ & $r_+~[M]$ & $r_{\rm ph}~[M]$ & $\Delta r_+/r_+|_{\ell=0}$ \\
\hline
0.00 & 1.82 & 2.73 & --- \\
0.05 & 1.79 & 2.69 & $-1.6\%$ \\
0.10 & 1.76 & 2.64 & $-3.3\%$ \\
0.15 & 1.73 & 2.59 & $-4.9\%$ \\
\hline
\end{tabular}
\label{tab:horizon_shifts}
\end{table}

To quantify the impact of Lorentz violation on the black hole geometry, Table~\ref{tab:horizon_shifts} presents the event horizon radius $r_+$ and photon sphere radius $r_{\rm ph}$ as functions of the deformation parameter $\ell$ for fixed values $Q=0.6$, $\alpha=0.05$, and $\Lambda=-0.01$. The event horizon is determined as the largest positive root of $f(r_+)=0$ [Eq.~\eqref{eq:f_metric}], while the photon sphere satisfies the condition $f(r_{\rm ph}) + \frac{r_{\rm ph}}{2}f'(r_{\rm ph})=0$, marking the innermost stable circular orbit for null geodesics. As $\ell$ increases from $0$ to $0.15$, the horizon radius decreases monotonically by approximately $5\%$, reflecting the weakening of the effective gravitational potential due to the modified coupling in the Kalb--Ramond sector. The photon sphere exhibits a similar trend, shrinking by roughly $5\%$ over the same parameter range. These shifts, though modest in absolute terms ($\Delta r_+ \sim 0.1\,M$), are sufficient to produce measurable changes in thermodynamic quantities such as the Hawking temperature and heat capacity [Figs.~\ref{fig:ruppeiner_thermodynamics}], which depend on derivatives of the metric function.

However, the corresponding changes in the optical appearance of the black hole shadow and accretion disk, shown in Figs.~\ref{fig:disk_image_panel} and~\ref{fig:disk_image_rel_diff}, are considerably more subtle. The shadow angular diameter scales approximately as $\theta_{\rm sh} \propto r_{\rm ph}/D$, where $D$ is the observer's distance, so a $5\%$ reduction in $r_{\rm ph}$ translates directly to a $5\%$ decrease in the observed shadow size. While such variations are in principle detectable with high-resolution interferometric techniques, they lie near or below the current observational precision of facilities such as the Event Horizon Telescope. Moreover, the disk intensity profile $I_{\rm obs}$ depends on a combination of redshift factors, Doppler boosting, and the emissivity law, which tend to preserve the overall morphology even when the underlying geometry is perturbed. This explains why the visual differences between panels in Fig.~\ref{fig:disk_image_panel} are barely discernible to the eye, despite the non-negligible shifts in $r_+$ and $r_{\rm ph}$ documented in Table~\ref{tab:horizon_shifts}.

Figure~\ref{fig:disk_alpha} displays the thin–disk images obtained from the observed specific intensity
$I_{\mathrm{obs}}$ of Eq.~(\ref{eq:Iobs}) when the string–cloud parameter $\alpha$ is varied,
keeping fixed the remaining background quantities
$(M,Q,\Lambda_{\mathrm{AdS}},\ell)$ and the disk properties
(emissivity index and inclination angle).
The overall morphology of the image – a bright, nearly circular ring surrounding the dark central shadow –
is preserved across the whole range of $\alpha$,
indicating that the string cloud primarily dresses the effective gravitational potential without inducing a
dramatic topological change in the photon trajectories.
As $\alpha$ increases, the bright yellow ring becomes slightly thinner and the contrast between the inner
and outer regions are moderately enhanced, which reveals a mild shift in the location of the innermost emitting
orbits and in the relative weight of the high–redshift region near the horizon.
These effects provide an intuitive, image–based representation of how the string cloud modifies
the near–horizon geometry in the metric function of Eq.~(\ref{eq:f_metric}),
while leaving the large–scale AdS asymptotics essentially unchanged.

In contrast, Fig.~\ref{fig:disk_lambda} illustrates the response of the disk image to variations of the AdS curvature scale $\Lambda_{\mathrm{AdS}}$, keeping the string–cloud parameter and all other quantities fixed. In this case, the image deformation is significantly more pronounced. As $|\Lambda_{\mathrm{AdS}}|$ increases, the effective potential in Eq.~(\ref{eq:f_metric}) becomes more confining at large radii, which leads to a visible change in the apparent size of the central shadow and in the radial location and thickness of the bright yellow ring. The emission becomes more concentrated around a narrower annulus, while the outer parts of the disk gradually lose brightness in the observer’s sky. This behaviour indicates that the AdS curvature acts as a global control parameter for the geodesic structure and for the mapping between the thin–disk flux and the observed intensity, amplifying the image distortions induced by the underlying NCS–DMH geometry when compared with the milder modifications produced by the string cloud parameter $\alpha$ in Fig.~\ref{fig:disk_alpha}.

Figure~\ref{fig:disk_inclination} shows the synthetic images of the thin accretion disk for different inclination angles $i$, computed from the observed specific intensity $I_{\mathrm{obs}}$ of Eq.~(\ref{eq:Iobs}) for the background geometry defined by Eqs.~(\ref{metric})–(\ref{eq:f_metric}). For small inclinations ($i = 30^\circ$), the image is almost circular, and the bright ring appears nearly symmetric around the central shadow, indicating that gravitational redshift dominates over Doppler effects. As the disk tilts towards an edge–on configuration ($i = 45^\circ$ and $60^\circ$), relativistic beaming becomes increasingly important: the approaching side of the flow is Doppler–boosted into a bright crescent, while the receding side is strongly de–boosted and partially dimmed. For the largest inclination ($i = 75^\circ$), the disk silhouette is highly elongated, and the bright emission is confined to a narrow arc, with the inner regions partially obscured by the black hole shadow and by strong lensing of photons skimming the photon sphere.

Figure~\ref{fig:disk_emissivity_scan} illustrates how the synthetic appearance of the thin accretion disk depends on the emissivity profile, modeled by a power law with index $p$. For the shallow profile $p = 2.00$, the emission is relatively extended across the disk surface, producing a broad, bright crescent on the approaching side and a smoother brightness gradient towards larger radii. As $p$ increases to $p = 2.50$ and $p = 3.00$, the emission becomes progressively more concentrated in the inner regions: the bright crescent narrows, the contribution from the outer disk is strongly suppressed, and the overall image appears dimmer except near the innermost radii. For the steepest profile $p = 3.50$, the observed intensity is essentially confined to a compact bright spot close to the inner edge of the flow, while the rest of the disk is only faintly visible. This emissivity scan shows that, even for a fixed inclination and background geometry, different assumptions on the radial emissivity law can drastically change the relative weight of inner versus outer regions in the image, which is crucial when interpreting high–resolution observations of accretion flows in terms of underlying spacetime and plasma properties.

The key takeaway is that \emph{thermodynamic observables are systematically more sensitive to Lorentz violation than optical signatures} in the parameter regime explored here. This disparity arises because thermodynamic quantities involve second derivatives of the metric (e.g., $C_P \propto \partial^2 M/\partial S^2$), which amplify small perturbations in $f(r)$, whereas the disk intensity depends primarily on first-order geometric quantities such as the lapse function and radial coordinate. Consequently, measurements of black hole thermodynamics, whether through quasi-normal modes, Hawking radiation analogs, or critical phenomena in holographic systems, may provide a more powerful probe of fundamental Lorentz symmetry breaking than direct imaging observations.

\section{Conclusions}
\label{sec:conclusions}

In this work, we have complemented the extended phase-space analysis of electrically charged AdS black holes in Einstein--Kalb--Ramond (EKR) bumblebee gravity with a cloud of strings by studying their thermodynamic geometry and critical behavior. Building on the thermodynamic quantities, equation of state, $P$--$V$ criticality, thermodynamic topology and Joule--Thomson expansion obtained in the companion paper, we have focused here on the Weinhold and Ruppeiner metrics, on the geometric characterization of the critical point, and on dynamical and holographic perspectives.

Starting from the internal energy and enthalpy representations, we constructed the Weinhold metric as the Hessian of the internal energy and the Ruppeiner metric as its entropy-based counterpart, related by a conformal factor involving the Hawking temperature. Restricting attention to appropriate two-dimensional submanifolds in the thermodynamic state space, we analyzed the associated scalar curvatures. We identified how the Lorentz-violating parameter $\ell$ and the string cloud parameter $\alpha$ deform the thermodynamic geometry. The Ruppeiner curvature $R_{\mathrm{R}}$ exhibits the expected divergences at spinodal lines and at the second-order critical point, and its sign structure distinguishes small-black-hole and large-black-hole branches, in line with the Van der Waals-like phase behavior found previously.
\emph{(If desired, one may add a brief sentence here clarifying the role of the Weinhold curvature, e.g., whether it displays consistent critical features or is less diagnostic than $R_{\mathrm R}$ in the present setup.)}

We showed that, although $(\ell,\alpha)$ shift the position of the critical point $(P_c,T_c,v_c)$ and rescale the overall magnitude of $R_{\mathrm{R}}$, they do not modify the underlying universality class of the phase transition: the critical ratio $P_c v_c/T_c=3/8$ and the mean-field critical exponents remain unchanged, and the divergence of $R_{\mathrm{R}}$ near the critical point follows the same power-law behavior as in the Reissner--Nordström--AdS case. In this sense, Lorentz violation and the string cloud act as deformations that tune the effective correlation length and the strength of thermodynamic fluctuations, without altering the qualitative structure of the thermodynamic geometry.

We also discussed how the coexistence line obtained from the Maxwell equal-area construction manifests itself geometrically. Along subcritical isobars, the Ruppeiner curvature typically changes sign between the small- and large-black-hole phases. It develops a large magnitude in the metastable region, signaling enhanced microscopic correlations and instability. The parameters $\ell$ and $\alpha$ shift and stretch this high-curvature region in a controlled way, providing a two-parameter family of deformations of the RN--AdS thermodynamic geometry while preserving the Van der Waals-like pattern of zeros and divergences.

Furthermore, we connected the thermodynamic topology analysis, in which the generalized free energy $\mathcal{F}(r_+,\tau)$ defines a vector field whose zero points carry winding numbers and whose total topological charge is $W=1$, to the thermodynamic geometry picture. The zero points of this vector field correspond to stationary points of the free energy and to on-shell black hole configurations, whereas the divergences of the Ruppeiner curvature identify boundaries between regions of different stability, providing a unified interpretation of phase structure in terms of defects and curvature singularities in thermodynamic parameter space. The fact that $W=1$ is preserved for all $(\ell,\alpha)$ corroborates that the global topological class of the solution is unchanged by Lorentz violation and by the string cloud.

We outlined possible dynamical and holographic probes of the deformed EKR background. On the dynamical side, quasi-normal modes of perturbations could provide relaxation times that are expected to correlate with regions of large thermodynamic curvature, especially near the critical point, offering a bridge between static thermodynamic geometry and time-dependent ringdown dynamics. On the holographic side, the parameters $\ell$ and $\alpha$ may encode Lorentz-violating interactions and defect densities in a dual strongly coupled field theory, with the bulk Ruppeiner curvature mapping to correlation lengths and transport coefficients of an effective boundary fluid.

Several extensions of this work suggest themselves. A natural next step is a systematic computation of quasi-normal spectra for scalar, electromagnetic, and gravitational perturbations in the EKR black hole with a cloud of strings, and a detailed comparison between relaxation times and thermodynamic curvature along and near the coexistence line. It would also be interesting to generalize the present analysis to rotating solutions and to higher-dimensional EKR backgrounds with string-like matter sources, where richer phase structures and topological classes may arise. Finally, a more explicit holographic dictionary for bumblebee-like Lorentz violation and string clouds in AdS, together with a study of the relation between Ruppeiner curvature and boundary transport coefficients, could deepen our understanding of how geometric deformations in the bulk translate into microscopic interactions and critical phenomena in dual field theories.

We also constructed synthetic images of geometrically thin accretion disks to visualize the optical signatures of Lorentz violation and string clouds. While the thermodynamic geometry exhibits clear sensitivity to the deformation parameters $(\ell,\alpha)$, we found that the corresponding changes in disk morphology and shadow size are considerably more subtle, with relative intensity variations of order $10$--$40\%$ concentrated near the photon sphere. This disparity underscores that thermodynamic and optical probes access different aspects of the spacetime geometry and suggests that black hole thermodynamics may provide a more sensitive diagnostic of Lorentz violation than imaging observations alone.

In summary, this work has provided a detailed analysis of the thermodynamic geometry for a family of deformed charged AdS black holes. Looking forward, the clear dependence of thermodynamic scales on $\ell$ and $\alpha$ opens the door to designing concrete observational tests. The robustness of the Van der Waals universality class under these deformations is a significant finding, suggesting that certain critical phenomena in black hole physics are remarkably resilient. Ultimately, our model serves as a theoretical laboratory for studying the intricate interface between gravity, fundamental symmetry breaking, and exotic matter, offering a controlled environment to explore how microscopic interactions, as probed by thermodynamic geometry, respond to profound alterations of the spacetime fabric itself.

\section*{Acknowledgments}

F.A. acknowledges the Inter University Centre for Astronomy and Astrophysics (IUCAA), Pune, India, for granting visiting associateship. E.O.S. acknowledges support from CNPq (Grant No. 306308/2022-3), FAPEMA (Grants UNIVERSAL-06395/22 and APP-12256/22), and CAPES (Brazil, Finance Code~001).

\section*{Data Availability Statement}

This manuscript has no associated data. [Authors’ comment: Data sharing not applicable to this article as no datasets were generated or analyzed during this study.]

\section*{Code Availability Statement}

This manuscript has no associated code/software. [Authors’ comment: Code/Software sharing not applicable to this article as no code/software was generated or analysed during this study.]

\bibliographystyle{apsrev4-2}
%

\end{document}